\documentclass[11pt]{revtex4}
\usepackage{amsmath}
\usepackage{latexsym}
\usepackage{float}
\usepackage{amssymb}
\usepackage{graphicx}
\usepackage{graphics}

\begin{document}

 \title{Interplay of order-disorder phenomena and diffusion in
     rigid binary alloys: Monte Carlo simulations of the
        two-dimensional ABV model}

 \author{A. De Virgiliis and Kurt Binder \\
    \textit{Institut f\"ur Physik, 
           Johannes-Gutenberg-Universit\"at Mainz,
           55099 Mainz, Germany}}

\begin{abstract}
Transport phenomena are studied for a binary (AB) alloy
on a rigid square lattice with nearest-neighbor attraction between
unlike particles, assuming a small concentration $c_v$ of
vacancies $V$ being present, to which $A(B)$ particles can jump
with rates $\Gamma _A (\Gamma _B)$ in the case where the nearest
neighbor attractive energy $\epsilon_{AB}$ is negligible in
comparison with the thermal energy $k_BT$ in the system. This
model exhibits a continuous order-disorder transition for
concentrations $c_A,c_B=1-c_A-c_V$ in the range
$c_{A,1}^{crit}\leq c_A \leq c_{A,2}^{crit}$, with
$c_{A,1}^{crit}=(1-m^*-c_V)/2$, $c_{A,2}^{crit} =(1+m^*-c_V)/2$,
$m* \approx 0.25$, the maximum critical temperature occurring for
$c*=c_A=c_B=(1-c_V)/2$, i.e. $m^*=0$. This phase transition
belongs to the $d=2$ Ising universality class, demonstrated by a
finite size scaling analysis. From a study of mean-square
displacements of tagged particles, self-diffusion coefficients are
deduced, while applying chemical potential gradients allow the
estimation of Onsager coefficients. Analyzing finally the decay 
with time of sinusoidal concentration variations that were prepared 
as initial condition, also the interdiffusion coefficient is
obtained as function of concentration and temperature. As in the
random alloy case (i.e., a noninteracting ABV-model) no simple
relation between self-diffusion and interdiffusion is found.
Unlike this model mean field theory cannot describe interdiffusion, 
however, even if the necessary Onsager coefficients are estimated 
via simulation.
\end{abstract}

\maketitle

\section{INTRODUCTION}
Understanding of atomic transport in multicomponent solids has
been a longstanding challenge \cite{1}-\cite{13}. In particular,
the problem of interdiffusion in binary metallic alloys (as well
as other types of mixed crystals) is very intricate: there is a
delicate interplay between kinetic aspects that have a complicated
energetics (such as jump rates of the various kinds of atoms to
available vacant sites) and effects due to non-random arrangement
of these atoms on the lattice sites (a problem which needs to be
considered in the framework of statistical thermodynamics
\cite{14,15,16}). Even the simplistic limiting case of perfectly
random occupation of the sites of a rigid perfect lattice by two
atomic species ($A,B$) and a small fraction of vacancies ($V$), where
one assumes constant jump rates $\Gamma_A,\Gamma_B$ of the two
types of atoms to the vacant sites (i.e. jump rates that do not
depend on the occupation of the sites surrounding the vacant
sites), is highly nontrivial \cite{17}. One finds that neither the
self-diffusion coefficients $D_A, D_B$ nor the interdiffusion
coefficient $D_{int}$ can be analytically reliably predicted,
given $\Gamma_A, \Gamma_B$ and the average concentration
$c_A,c_B$; nor does a simple relation between $D_A, D_B$ and
$D_{int}$ exist \cite{17}.

Recently, attention has been focused on this problem because of
several fascinating developments: (i) Progress with the electronic
structure calculations of vacancy formation energies, jump rates
etc. as well as better understanding of short range order
parameters in alloys puts the ``first-principles'' calculation of
interdiffusion and self-diffusion coefficients in ordered solid
alloys such as $Al_{(1-x)}Li_x$ within reach \cite{13}. (ii)
Progress with the atom-tracking scanning tunnelling microscopy
observation of atomic motions in two-dimensional surface alloys
such as In atoms in Cu(001) surfaces \cite{10} or Pd atoms in
Cu(001) surfaces \cite{11} has provided
compelling direct evidence for the operation of vacancy mediated
surface diffusion. This is a nontrivial result, since competing
mechanisms (surface atoms leave the topmost atomic layer to become
adatoms on top of this layer \cite{18}, or direct exchange between
neighboring surface atoms, ``assisted'' by the free space above
the topmost monolayer of atoms at the crystal surface) can not be
ruled out a priori. Of course, this finding enforces the
hypothesis that vacancy mechanism dominates self- and
interdiffusion processes in crystal lattices in the bulk
\cite{1,2,3,4,5,6,7,8,9}.

In the present work we try to contribute to this problem,
emphasizing the statistical mechanics approach by considering
again a rigid lattice model but allowing for interactions causing
a nontrivial long range order (or, at least, short range order)
between the atoms in the system. We are not addressing a specific
material, but rather try to elucidate the generic phenomena caused
by the interplay of local correlations in the occupancy of lattice
sites and the disparities in the jump rates $\Gamma_A$ and
$\Gamma_B$ of the two species. Thus, our model is close in spirit
to the work in Ref. \cite{17} and employs a related Monte Carlo 
simulation methodology \cite{19}. Unlike \cite{17}, the present 
model does include a nearest-neighbor attraction between unlike 
neighbors, and thus nontrivial static order-disorder phenomena occur. 
As expected, we shall demonstrate that the resulting correlations in
the occupancy of the lattice sites have a drastic effect on the
transport phenomena, and hence cannot be neglected when one tries
to interpret real data. We also emphasize that these correlation
phenomena need a treatment beyond mean field level. We point out
this fact, because sometimes a first principle electronic
structure calculation is combined with statistical mechanics of
mean field type or the cluster variation method \cite{16}, and
such approximations then clearly spoil the desirable rigor.
We also note that similar models as studied here have been frequently
used to study domain growth in alloys that are quenched from the high
temperature phase to a temperature below the order-disorder transition
temperature \cite{ecm}.

In Section II we describe our (two-dimensional) model. We restrict
the present work to two-dimensional systems, since recently there
has been great interest in two-dimensional alloys \cite{20}, and
we hope that extensions of our modelling can make contact with
corresponding experiments. In Section III we summarize our
simulation methodology, while Section IV briefly reviews some
pertinent theoretical concepts and approximations we wish to test.
Section V describes our numerical results, while Section VI
summarizes our conclusions, and gives an outlook to future work.

\section{THE MODEL AND ITS STATIC PROPERTIES}

\begin{figure}
  \begin{center}
     \scalebox{0.25}{\includegraphics{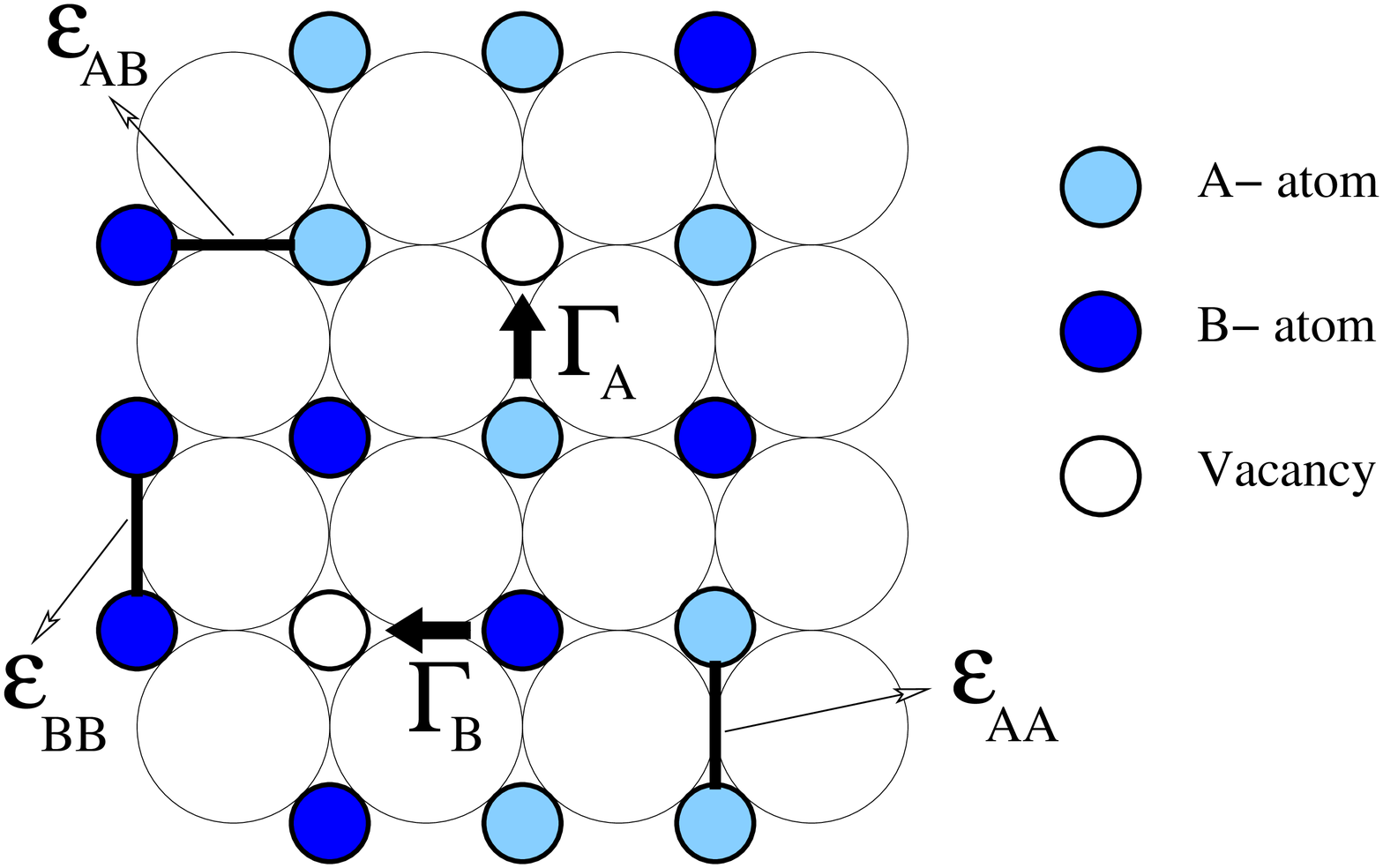}}
  \end{center}
  \caption{
     Schematic view of the $(100)$ surface of a substrate (shown as large open
     circles), whose periodic potential provides a square lattice of prefered
     adsorption sites (which here are assumed in the center of the square formed
     by the substrate atoms). $A$-atoms are shown as black circles, $B$-atoms
     are shown as grey circles, and vacancies ($V$) are shown as empty circles.
     The energies of the nearest-neighbor interactions between different kind of
     atoms (indicated by thick lines) are labeled by $\epsilon_{AA}$, 
      $\epsilon_{BB}$ and $\epsilon_{AB}$ respectively. The simple choice
     $\epsilon_{AB} \equiv \epsilon$, $\epsilon_{AA} = \epsilon_{BB} \equiv 0$
      is taken througout. This means that $A-$atoms prefer $B-$atoms as nearest
     neighbors, but it does not matter whether its nearest neighbors are also
     $A-$atoms or vacancies, respectively.  
     The jump rates for $A-V$ and $B-V$ exchanges are labeled by $\Gamma_A$ and
     $\Gamma_B$, respectively. For simplicity, the $B-$atoms are considered as
     the faster particles ($\Gamma_B \equiv 1$), and $\Gamma_A < \Gamma_B $.
     }   \label{fig1}
\end{figure}

Having in mind the application of our work to two-dimensional
surface alloys, we assume a perfect square lattice of adsorption
sites (Fig.~\ref{fig1}). These adsorption sites can either be
taken by an A-atom, a B-atom, or a vacancy. Therefore this model
traditionally is also referred to as the ABV model \cite{17,21}.
It can also be viewed as an extension of simple lattice gas
models, where diffusion of a single species (A) occurs by hopping
to vacant sites, to two components. Diffusion in lattice gases
with a single species has been extensively studied
\cite{5,22,23,24,25,26,27,28,29,30}, but diffusion in a
two-component lattice gas so far has been thoroughly examined only
in the noninteracting case \cite{17}. Here we restrict attention
to a model with strictly pairwise interactions between nearest
neighbors only, which we denote as $\epsilon_{AA}, \epsilon_{AB}$
and $\epsilon_{BB}$ pairs. However, in general one can consider 
also energy parameters between pairs of lattice sites involving
one ($\epsilon_{AV}$, $\epsilon_{BV}$) or two ($\epsilon_{VV}$) 
vacancies, but here we do not consider the $ABV$ model in full
generality, but only the special case 
$\epsilon_{AV} = \epsilon_{BV} = \epsilon_{VV} = 0$, although 
from first principles electronic structure calculations there is 
evidence that nonzero $\epsilon_{AV}, \epsilon_{BV}, \epsilon_{VV}$ 
may occur \cite{bes}.
While all these parameters affect the diffusion behavior of the 
model, actually only a subset of them controls the static behavior.
With respect to static properties of this model, the well-known
transcription to the spin$-1$ Blume-Emery-Griffiths model shows
(see e.g. \cite{ecm}), that for constant concentrations only three 
interaction parameters would be needed.
Note that although there are three concentration variables, 
$c_A,c_B,$ and $c_V$, due to the constraint $c_A+c_B+c_V=1$ only 
two of them are independent.
Actually, the physically most interesting case is the limit $c_V
\rightarrow 0$, since in thermal equilibrium the concentration of
vacancies is very small. In the noninteracting case \cite{17}, it
was found that many aspects of this limiting behavior $c_V
\rightarrow 0$ are already reproduced if the vacancy concentration
is of the order of a few percent only, e.g. $c_V=0.04$, and in
fact we shall adopt this choice in the case of the present
simulations. Also for the interacting case the limit $c_V
\rightarrow 0$ greatly simplifies matters, since then, with respect
to static properties, we have to consider only a single energy 
parameter $\epsilon$, defined by

\begin{equation} \label{eq1}
   \epsilon \equiv \epsilon_{AB}-(\epsilon_{AA}+\epsilon_{BB})/2 .
\end{equation}

If $\epsilon < 0$, the model in thermal equilibrium will exhibit
ordering, while for $\epsilon > 0$, phase separation occurs
\cite{14,15,31}. In the case of a square lattice, the model in the
limit $c_V\rightarrow 0$ is equivalent to the two-dimensional
Ising model, for which some static properties of interest are
exactly known \cite{32,33,34}. In particular, for $c_A=c_B=1/2$
the critical temperature $T_c$ is known exactly, namely
\begin{equation} \label{eq2}
k_BT_c^{max}/|\epsilon| = [\ln(1+\sqrt{2})]^{-1} \approx 1.1345 .
\end{equation}
This is the maximum value of the critical temperature curve
$T_c(c_A)$ at which the order-disorder phase transition
occurs. According to the well-known Bragg-Williams mean-field
approximation, one would rather obtain $k_B T_c^{MF} / \epsilon = 2$
than the result implied by Eq.~(\ref{eq2}), $k_B T_c / \epsilon
\approx 1.1345$ \cite{35}. Here and in the following, the maximum
value $T_c(c_A=0.5)$ of the pure model without vacancies is simply
denoted as $T_c$. However, an even more important failure
of the mean field theory is the prediction that an order-disorder
transition from the disordered phase to a phase with long-range
checkerboard order occurs over the entire concentration range,
with $T_c(c_A \rightarrow 0) \rightarrow 0,\; T_c(c_A \rightarrow
1)\rightarrow 0$, see \cite{35} for a more detailed discussion of
mean field theory. As a matter of fact, long range order is only
possible for a much more restricted range of concentrations,
namely \cite{35} $0.375 \leq c_A \leq 0.625$ (note that the
pairwise character of the interactions implies a symmetry of the
phase diagram around the line $c_A=1/2$, in the limit $c_V
\rightarrow 0$ \cite{14,15,31}).

\begin{figure}
  \begin{center}
     \scalebox{0.5}{\includegraphics{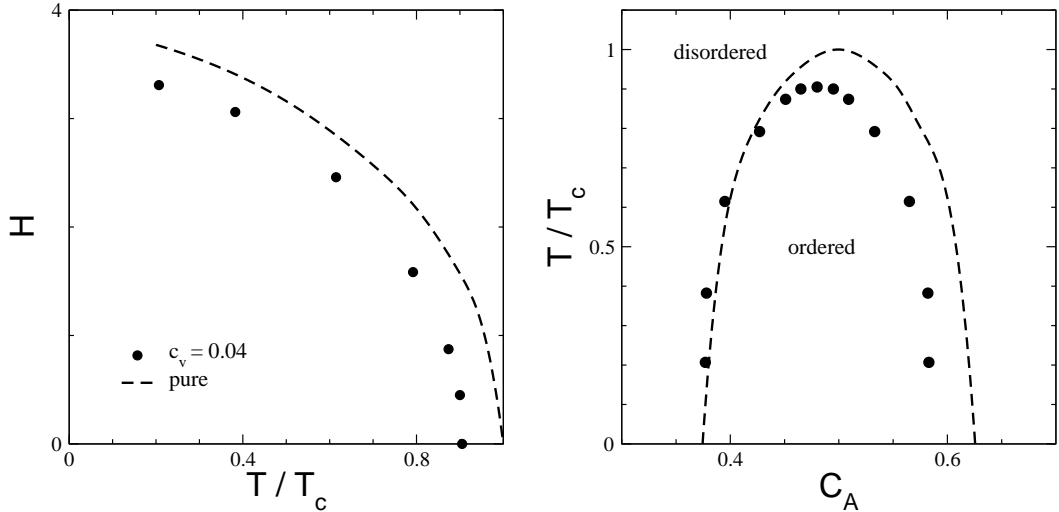}}
  \end{center}
  \caption{
       ($a$) Phase boundary for the order-disorder transition of the $ABV$ model 
       with $c_V = 0.04$. The phase boundary of the pure Ising antiferromagnet
        \cite{35} (equivalent to the case $c_V=0$) is also included for comparison, 
       as a dashed line. 
       ($b$) Critical curve $T$ vs. $c_A$, where $c_A = 1-c_V-c_B$. 
        The critical curve of the pure Ising antiferromagnet \cite{35} 
        is also included for comparison, as a dashed line. Temperature $T$ is
       always measured in units of the maximal critical temperature $T_c$ of
       the pure model (no vacancies, $c_V = 0$ and $c_A = c_B = 0.5)$, cf. text.
          }   \label{fig2}
\end{figure}

\begin{figure}
  \begin{center}
     \scalebox{0.5}{\includegraphics{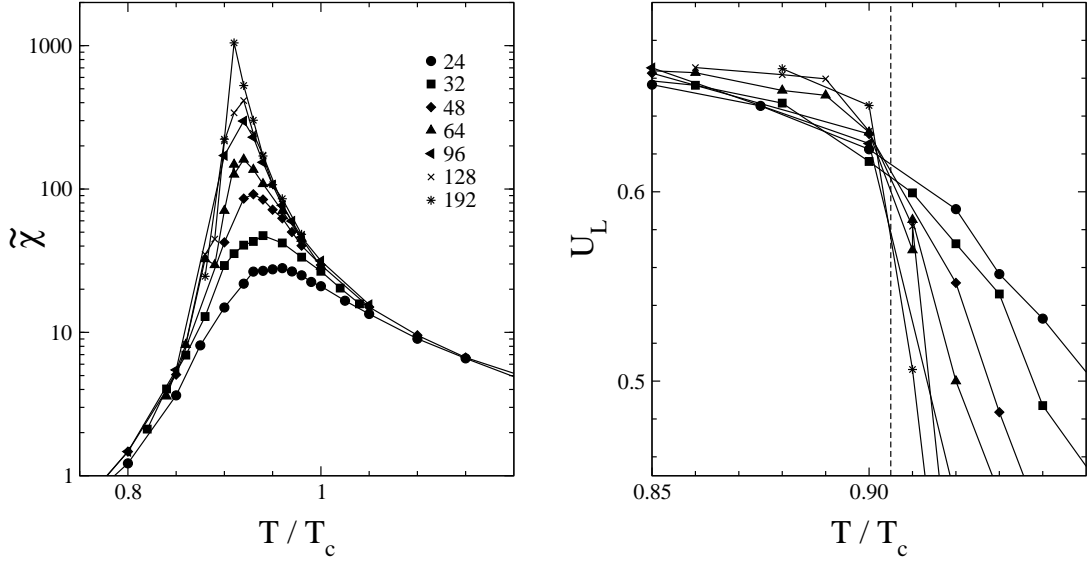}}
  \end{center}
  \caption{
       Dependence of the staggered susceptibility $\tilde{\chi}$ ($a$) and the
       fourth order cumulant $U_L$ ($b$) on the temperature, along the
       critical line $H=0$ corresponding to the critical concentration
       $c_A=0.48$. Several system sizes are considered, as indicated. 
       Here $T_c$ denotes the maximal critical temperature of the model 
       without vacancies ($H=0$ then corresponds to $c_A = 0.5$).
          }   \label{fig3}
\end{figure}

\begin{figure}
  \begin{center}
     \scalebox{0.5}{\includegraphics{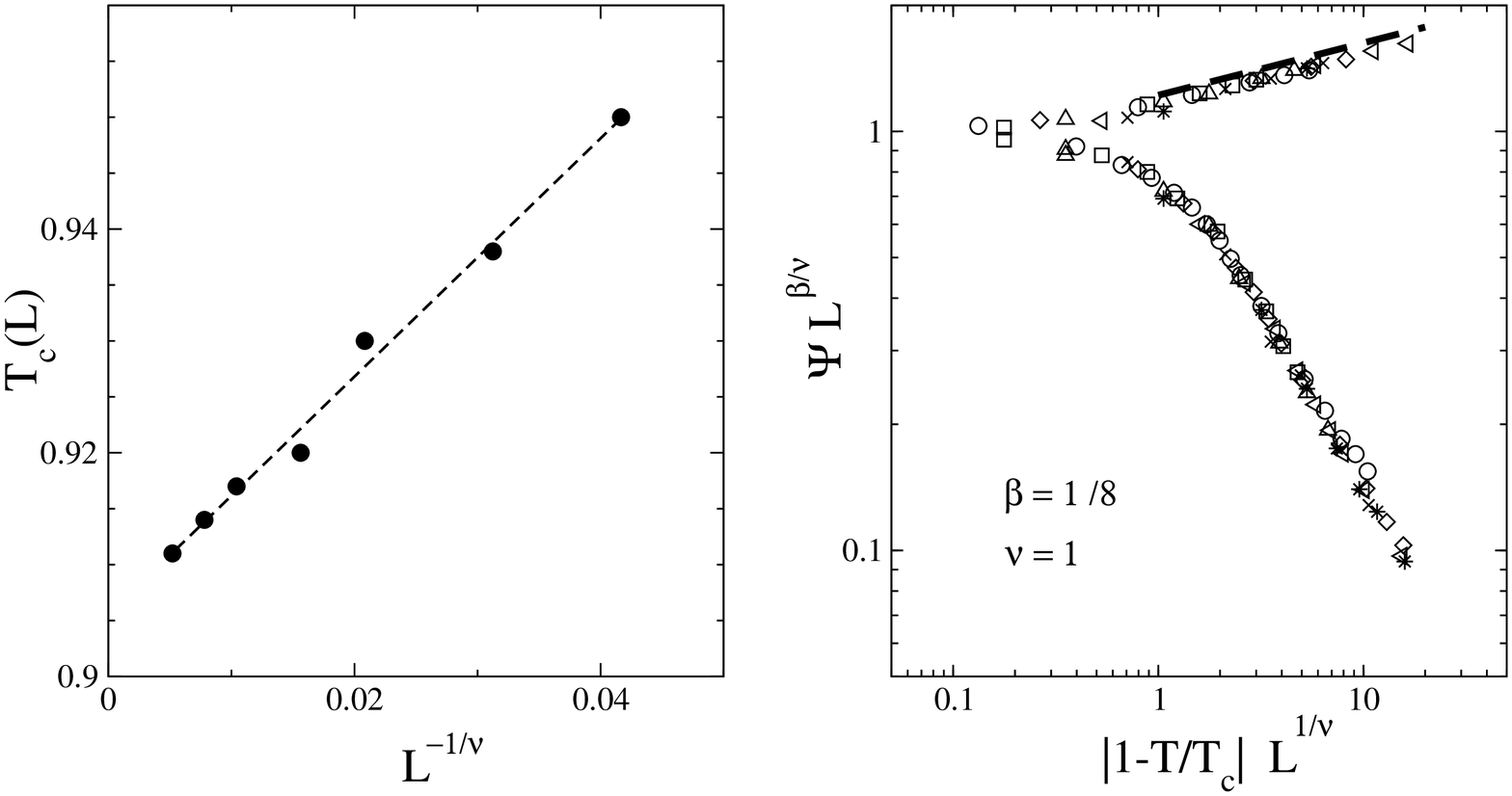}}
  \end{center}
  \caption{
     ($a$) Plot of the size-dependent critical temperature $T_c(L)$ (defined as
      the maximum of $\tilde{\chi}(T,L)$), in terms of the scaled variable $L^{1/\nu}$. 
      The critical Ising exponent $\nu=1$ is employed. The linear extrapolation
      to the thermodynamic limit, shown as a dashed line, provides an estimation
      of  $T_c(c_V=0.04) = 0.905(5)$ for the $ABV$ model with $c_V = 0.04$.
      ($b$) Scaling plot of the order parameter $\psi$. The estimated critical
      temperature and the Ising critical exponents $\nu=1$ and $\beta=1/8$ are 
      employed. 
          }   \label{fig4}
\end{figure}

If we work with a small but nonzero concentration of vacancies
$c_V$, the maximum critical temperature no longer occurs at
$c_A = c_B = 1/2$, but rather at $c^* = c_A = c_B=(1-c_V)/2$, 
and the phase diagram is in this case symmetric around this 
concentration $c^*$. Apart from this statement, there are no 
longer any exact results available, but it is fairly straightforward 
to obtain the phase diagram from standard Monte Carlo methods \cite{19} 
with an accuracy that is sufficient for our purposes. Fig.\ref{fig2} 
shows our estimates of the phase boundary for $c_V=0.04$, in comparison
with previous results for $c_V=0$. As has been well documented in
the literature \cite{19,31,35}, such phase diagrams are
conveniently mapped out by transforming the model to a magnetic
Ising spin model (representing the cases that lattice site $i$ is
taken by an $A-$atom by spin up, $B-$atom by spin down, respectively)
and considering the transition from the paramagnetic to the
antiferromagnetic phase for various magnetic fields $H$
($2\;H = \mu_A - \mu_B$, if $\epsilon_{AA} = \epsilon_{BB}$, 
and with $\mu_A$ and $\mu_B$ being the chemical potentials of
$A$ and $B-$ particles, respectively).
Estimating then the magnetization $m_c(H)=m(H,T=T_c(H))$ at the
phase boundary, one then obtains the corresponding critical
concentrations from
\begin{equation} \label{eq3}
      c^{crit}_A(T)=(1\pm m_c(T)-c_V)/2
\end{equation}
Fig.~\ref{fig2} shows that for $c_V=0.04$ the maximum critical
temperature occurs for $T_c(c_V=0.04)/T_c\approx 0.905$, and for
$T \rightarrow 0$ the phase boundary ends at the concentrations
$c_{A,1}^{crit} \approx 0.375,\; c_{A,2}^{crit} \approx 0.585$. As
it should be, the phase diagram is symmetric around
$c_{A,max}^{crit} =(1-c_V)/2=0.48$. The analysis indicates that
the order-disorder transition stays second order throughout, also
in the presence of this small vacancy concentration. Although it
is clear that a vacancy concentration of $c_V=0.04$ does have some
clearly visible effects, in comparison to the model with 
$c_V \rightarrow 0$, these changes do not affect the qualitative 
character of the phase behavior, but cause only minor
modifications of quantitative details. For obtaining accurate
results on the dynamic behavior of the model with a modest amount
of computing time, working with sufficiently many vacancies on the
lattice is mandatory. Note that for the diffusion studies we use a
lattice of linear dimensions $L=1024$, while the static phase
diagram was extracted from a standard finite size scaling analysis
\cite{19}, see Figs.~\ref{fig3}, \ref{fig4} for an example, using
sizes $24 \leq L \leq 192$. Periodic boundary conditions are
applied throughout. The static quantities that were analyzed in
order to obtain the phase boundary are the antiferromagnetic order
parameter $\Psi$ (we refer here to the transformation of the model
to the Ising spin representation again)
\begin{equation} \label{eq4}
   \Psi = \langle |\phi| \rangle, \;\;\;  \phi = L^{-2} \sum \limits ^L
   _{k=1} \sum \limits _{\ell =1}^L (-1)^{k+\ell} S_{k, \ell} \,\, ,
\end{equation}
where $k,\ell$ label the lattice sites in $x-$ and $y-$ direction,
respectively. Similarly, the magnetization $m$ is given by averaging
all the spins without a phase factor
\begin{equation} \label{eq5}
  m=\langle M \rangle , \quad M=L^{-2} \sum \limits _{k=1}^L \sum 
  \limits _{\ell =1}^L S_{k,\ell} \,\, ,
\end{equation}
and the susceptibility $\chi$ and staggered susceptibility
$\tilde{\chi}$ are obtained from the standard fluctuation relations
\begin{equation} \label{eq6}
  \chi = L^2(\langle M ^2\rangle - \langle M \rangle ^2)/k_BT \;, 
\end{equation}
\begin{equation} \label{eq7}
  \tilde{\chi} = L^2(\langle \phi ^2 \rangle - \langle |\phi| \rangle 
               ^2)/k_BT \,\, .
\end{equation}
Note that in a finite system in the absence of symetry-breaking fields
one needs to work with the average of the absolute value 
$\langle |\phi| \rangle$ rather than $\langle \phi \rangle$ in order 
to have a meaningful order parameter \cite{19}.

A further quantity useful for finding the location of the transition 
is the fourth order cumulant of the order parameter \cite{36}
\begin{equation} \label{eq8}
   U_L=1-\langle \phi ^4\rangle /[3 \langle \phi ^2\rangle ^2] \, ,
\end{equation}
since the critical temperature can be found from the intersection
of the cumulants plotted versus temperature for different lattice
sizes. For the two-dimensional Ising universality class, this
intersection should occur for a value $U^* \approx 0.6107$
\cite{37}.

Fig.~\ref{fig3} shows that this expectation is only rather roughly
fulfilled. To some extent this may be attributed to statistical
errors, but in addition probably for $c_V \neq 0 $ there are
somewhat larger corrections to finite size scaling than for the
``pure'' model (i.e., the model without vacancies). We have hence
estimated $T_c(c_V=0.04)$ alternatively from a plot of the
temperatures $T_c(L)$, where the maximum of $\tilde{\chi}(T,L)$
for finite $L$ occurs, versus the finite size scaling variable
$L^{-1/\nu}=L^{-1}$ (remember $\nu =1$ in the two-dimensional
Ising model \cite{34}), see Fig.~\ref{fig4}a. The quality of the
finite size scaling ``data collapse'' of the order parameter
(fig.~\ref{fig4}b) gives us confidence in the reliability of our
procedures.

We emphasize that the present paper concerns only the choice of the
symmetric case, $\epsilon_{AA} = \epsilon_{BB}$. While any asymmetry
between $A$ and $B$, leading to $\epsilon_{AA} \neq \epsilon_{BB}$,
has little effect on static properties for small $c_V$, the distribution
of the vacancies and their dynamics may get strongly affected by such
an asymmetry \cite{ecm,21}. 

Finally, we mention a static quantity that plays a role in discussing 
the self-diffusion coefficient of particles in lattice gas models, 
the so-called ``vacancy availability factor'' \cite{5,23,29}
\begin{equation} \label{eq9}
   V=c_V(1-\alpha_1)
\end{equation}
Here $\alpha_1$ is the standard Cowley-Warren short range order
parameter \cite{14,15,16,31} for the nearest neighbor shell of a
particle: $\alpha_1=0$ if there is a random occupation of the
lattice sites by any particles and vacancies (note that here we
are not concerned with short range order describing the non-random
occupation of $A$ versus $B$ particles on the lattice. Due to the
symmetry $\epsilon_{AA}=\epsilon_{BB}$, there is also no need to
consider separate vacancy availability factors for $A$ and $B$
particles). Actually, we expect that in the limit $c_V \rightarrow
0$ also $\alpha _1 \rightarrow 0$, and then $V=c_V$. Hence a
calculation of $\alpha _1$ can serve as a test whether the chosen
vacancy concentration is small enough in order to reproduce the
desired limit $c_V \rightarrow 0$.

\section{SIMULATION METHODOLOGY TO STUDY TRANSPORT PHENOMENA} \label{secIII}
The Monte Carlo simulations consist of an initial part, necessary
to equilibrate the system for the desired conditions, and a final
part, where the transport coefficients of interest are
``measured'' in the simulation. While in the case of the
completely random $ABV$ alloy studied in \cite{17} the generation of
an initial configuration is straightforward, this is not so here,
because depending on where the chosen state point $(T,c_A)$ is in
the phase diagram, Fig.~\ref{fig2}, we have long range order or
not. If the system in equilibrium is in a state where long range
order occurs, it is important to prepare the system in a monodomain 
sample: otherwise the presence of antiphase domain boundaries \cite{31}
might spoil the results. In particular, at very low temperature
interdiffusion could be strongly enhanced near such boundaries, in
comparison with the bulk. Although such effects are interesting in
their own right, they need separate study from bulk behavior, and
are out of consideration here.

Actually the best way to prepare the equilibrated initial
configurations, in cases where long range order is present, is the
use of the ``magnetic'' representation of the model as an Ising
antiferromagnet in a field $H$ (remember that $H$ physically
corresponds to the chemical potential difference between $A$ and $B$
particles \cite{31}). Recording the magnetization $m(T,H)$ as
function of the field, one can choose the field such that states
with the desired value of $m$ and hence $c_A=(1-c_V-m)/2$ result.
The initial spin configuration is that of a perfect
antiferromagnetic structure, from which a fraction $c_V$ of sites
chosen at random is removed. The Monte Carlo algorithm that was
used is the standard single spin flip Metropolis algorithm
\cite{19}, mixed with random exchanges of the vacancies with
randomly chosen neighbors. Note that during this equilibration
part the concentration $c_A$ is not strictly constant, but
slightly fluctuating: this lack of conservation of $c_A$ is
desirable, however, since ``hydrodynamic slowing down'' \cite{19}
of long wavelength concentration fluctuations would otherwise
hamper the equilibration of concentration correlations at long
distances.

\begin{figure}
  \begin{center}
     \scalebox{0.55}{\includegraphics{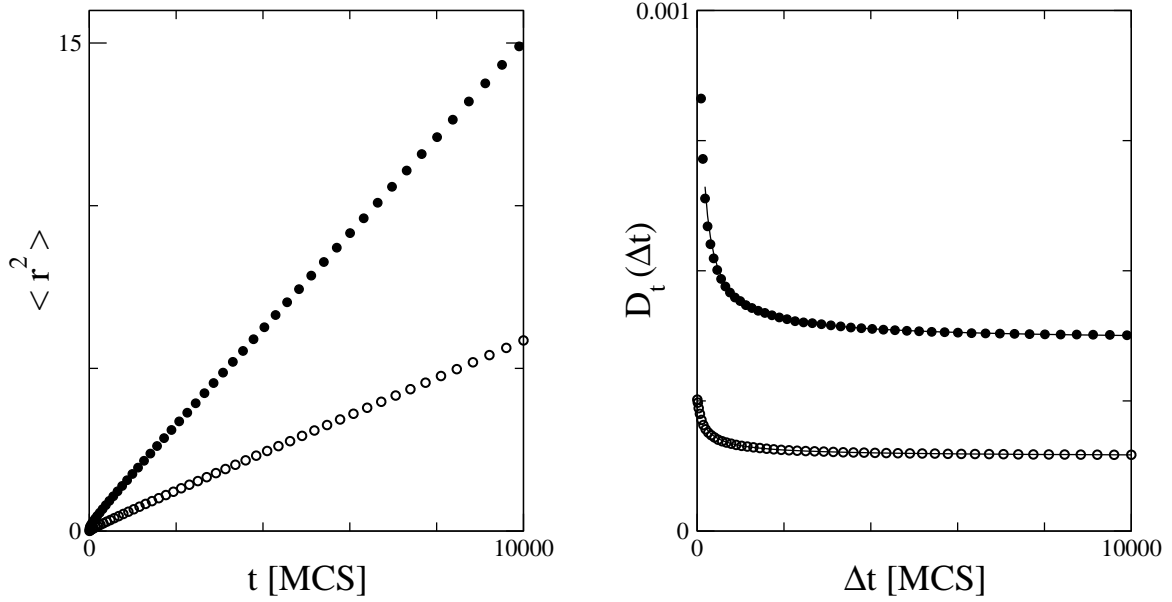}}
  \end{center}
  \caption{
     Determination of the tracer diffusion coefficients.
     ($a$) Mean square displacements of tagged $A$ particles (open dots) and
     $B$  particles (full dots)
     as a function of Monte Carlo steps per particle. The temperature is
     $T=1.2$ (in units of the Ising critical temperature), and the
     concentrations are $c_A=0.4$, $c_B=0.56$.
      The ratio of jump rates is $\Gamma_A /  \Gamma_B = 0.01$.
      ($b$) Estimates of the tracer diffusion coefficients as a function of the 
       time interval used. The lines represent the fits of the data after using
       Eq.~(\ref{eq11}).
          }   \label{fig5}
\end{figure}

In the final stage of the Monte Carlo runs, of course, the spin
flip Monte Carlo moves are shut off, since for the analysis of the
diffusion constants the concentrations $c_A,c_B=1-c_V-c_A$ need to
be strictly conserved. Most straightforward is the estimation of
the self-diffusion coefficients (also called ```tracer diffusion
coefficients'') $D_t$ of tagged $A$ and $B$ particles, since there one
simply can apply the Einstein relation
\begin{equation} \label{eq10}
  \langle r^2 \rangle = 2\,d\,D_t\,t, \;\;\;\;\;\;  t \rightarrow \infty, 
         \;\;\;\;\;\;   r = \vec{r}_i(t)-\vec{r}_i(0),
\end{equation}
$d$ being the dimensionality of the lattice ($d = 2$ here), and
$\vec{r}_i(t)$ being the position of the $i$-th particle at time $t$.
Fig.~\ref{fig5} illustrates the application of this method for a
typical example, in the case $\Gamma_A/\Gamma_B=0.01$, temperature
$T=1.2$ (in units of $T_c$, Eq.~\ref{eq2}), and concentrations
$c_A=0.40,c_B=0.56$, respectively. While the plot of $\langle
r^2 \rangle $ vs. $t$, for a total time of $t=10^4$ MCS, look at
first sight almost linear (Fig.~\ref{fig5}, left part), a closer
look reveals a slight but systematic decrease of the slope of the
$\langle r^2 \rangle$ vs. $t$ curve with increasing time. A similar
observation was already reported by Kehr et al. \cite{17}, who
attributed this decrease of slope to the presence of a logarithmic
correction.

Specifically, it was shown that in $d=2$ the estimate $D_{est}(\Delta t)$ 
of the tracer diffusion constants depend on the time interval $\Delta t$ 
of estimation as
\begin{equation} \label{eq11}
 D_{est}(\Delta t)= A + B \ln (\Delta t)/\Delta t + C/\Delta t \;,
\end{equation}
where $A,B,C$ are phenomenological constants. Therefore we have
analyzed $D_{est}(\Delta t)$ as a function of $\Delta t$ in the
present case (Fig.~\ref{fig5}, right part). We found rather
generally that there is a significant dependence of
$D_{est}(\Delta t)$ on $\Delta t$ for $\Delta t \leq 2.10^3$,
while for $\Delta t \geq 5.10^3$ the dependence on $\Delta t$ can
safely be neglected. A remarkable feature of the results also is
that the faster B particles exhibit (in the example shown in
Fig.~\ref{fig5}) a diffusion constant that is only about a factor
of three larger than the slower A particles, while the jump rate is
a factor of 100 larger. This fact already indicates that there is
no straightforward relation between the tracer diffusion constants
and the jump rates.

In the description of collective diffusion, the Onsager coefficients 
$\Lambda_{AA}$, $\Lambda_{AB}$, $\Lambda_{BA}$ and $\Lambda_{BB}$ play 
a central role, since they appear as coefficients in the linear relations 
between particle currents $\vec{j}_A,\vec{j}_B$ and the corresponding 
driving forces, the gradients of the potential differences between A (or B) 
particles and vacancies V, respectively \cite{17}:
\begin{equation} \label{eq12}
   \vec{j}_A=-(\Lambda_{AA}/k_BT)\nabla
   (\mu_A-\mu_V)-(\Lambda_{AB}/k_BT)\nabla(\mu_B-\mu_V),
\end{equation}
\begin{equation} \label{eq13}
   \vec{j}_B=-(\Lambda_{BA}/k_BT)\nabla
   (\mu_A-\mu_V)-(\Lambda_{BB}/k_BT)\nabla(\mu_B-\mu_V) \; .
\end{equation}
Note that due to the symmetry relation
\begin{equation} \label{eq14}
   \Lambda_{BA}=\Lambda_{AB}
\end{equation}
only three of these four Onsager coefficients are thought to be
independent. There is no simple relation between the two jump
rates $\Gamma_A,\Gamma_B$ (and temperature T and the
concentrations $c_A,c_B$) and these three Onsager coefficients
$\Lambda_{AA},\Lambda_{AB},\Lambda_{BB}$, of course. Hence it is a
task of the simulation to estimate these Onsager coefficients, and
it is well known \cite{17,22} that this can be done by applying a
force to the particles, which acts in the same way as a chemical
potential gradient. Due to the periodic boundary conditions,
particles that leave the box at one side will reenter at the
opposite one, and hence a chemical potential gradient causes a
steady state flux of particles through the simulation box in the
direction of this driving force. Care is needed in two respects:
\begin{itemize} 
 \item One must average long enough to make sure that slow
 transients after the imposition of the force have died out and
 steady-state conditions are actually reached.
  \item One must make sure that the applied force is small enough 
 so one works in the region where the response of the system to 
this force is strictly linear, as written in Eqs.~(\ref{eq12}), 
(\ref{eq13}), and nonlinear corrections can be completely neglected. 
\end{itemize}

\begin{figure}
  \begin{center}
     \scalebox{0.55}{\includegraphics{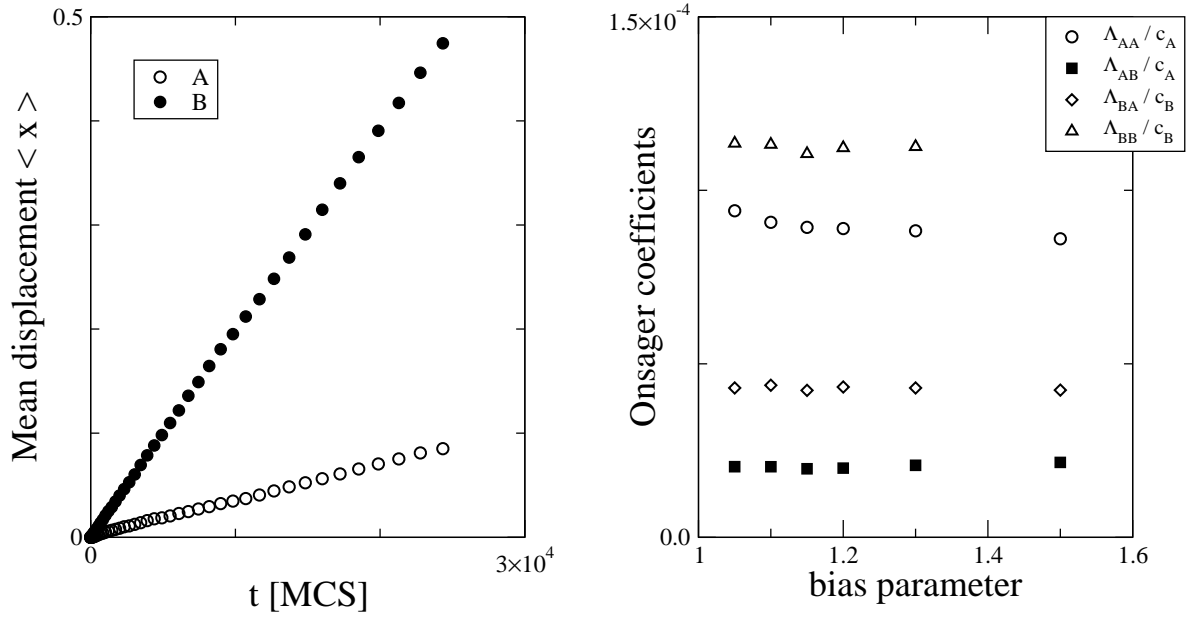}}
  \end{center}
  \caption{
     Determination of the Onsager coefficients.
     ($a$) Mean displacements along the $x-$ direction of $A$ and $B$
     particles as a function
     of Monte Carlo steps per particle. The temperature is $T=0.6$, and the 
      concentrations are $c_A=0.71$, $c_B=0.25$.
     The ratio of jump rates is $\Gamma_A /  \Gamma_B = 0.01$, and the bias parameter 
     is $b = 1.1$.
    ($b$) Estimates of the Onsager coefficients $\Lambda_{ij} / c_i$ by extrapolation
      to bias parameter $b=1$.
          }   \label{fig6}
\end{figure}

This method of estimating Onsager coefficients was pioneered by
Murch and Thorn \cite{22} for one-component lattice gases and
extended to random alloy models in \cite{17,38}. We refer the
reader to these papers for a more detailed justification and
discussion of this method. Following \cite{17} we implement this
force on species of particles $\gamma$ ($\gamma=A$ or $B$) by
taking the jump rates in the $x-$ direction as
\begin{equation} \label{eq15}
   \Gamma_x^{(\gamma)} = b\,\Gamma_\gamma, \;\; 
   \Gamma_{-x}^{(\gamma)} = b^{-1}\,\Gamma_\gamma, \;\; b>1
\end{equation}
while the jump rate in the $\pm y-$ directions remains
$\Gamma_\gamma$. If we would have a single particle (s.p.) only,
the mean velocity in the $+x-$ direction would be
$v_x^{s.p.}=\Gamma_\gamma(b-b^{-1})$, which should correspond to
$v_x^{s.p.}=(\Gamma_\gamma/k_BT)F_x$, $F_x$ being the force in
$x-$ direction, in the regime of linear response. Hence one
concludes that from the velocity of species $\alpha$ one can
deduce the Onsager coefficient $\Lambda_{\alpha \gamma}$ if a
force $F_x$ is exerted on species $\gamma$ via
\begin{equation}\label{eq16}
   v_x^{(\alpha)}/v^{s.p.}_x=(\Gamma_\alpha c_\alpha)^{-1}
                              \Lambda_{\alpha \gamma}.
\end{equation}
The application of this method is illustrated in Fig.~\ref{fig6}.
There the mean displacement $\langle x \rangle $ of $A-$ and
$B-$ particles is followed over $2.5 \times 10^4$ Monte Carlo steps
(MCS) per site, and a very good linearity of $\langle x \rangle$
vs. $t$ is observed (left part). In order to check for nonlinear
effects, the bias parameter $b$ is varied in the range $1.05 \leq b
\leq 1.5$, and the results are extrapolated to $b \rightarrow 1$.
(right part of Fig.~\ref{fig6}). Consistent with previous work on
the random $ABV$ model \cite{17}, nonlinear effects are rather
weak, and in this way we are able to estimate Onsager coefficients 
with a relative error of a few percent.

\begin{figure}
  \begin{center}
     \scalebox{0.55}{\includegraphics{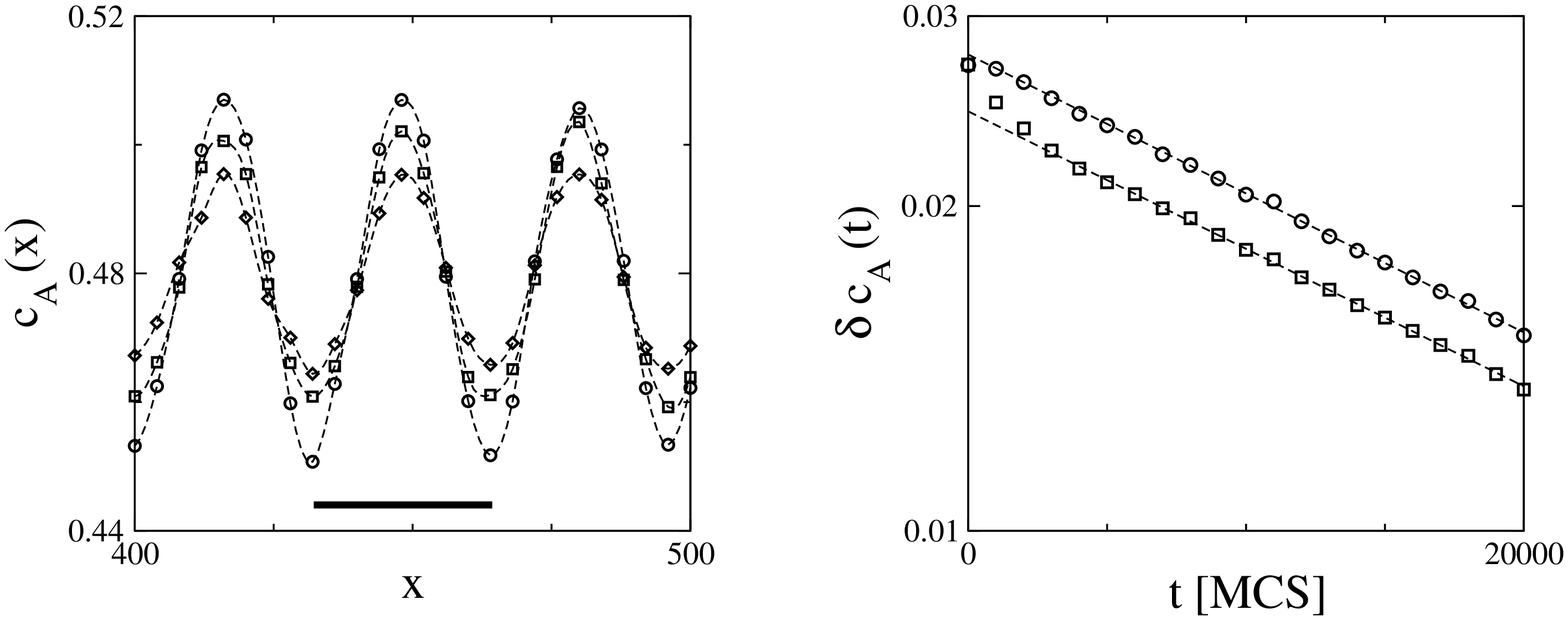}}
  \end{center}
  \caption{
     Determination of the interdiffusion coefficients.
     For $t < 0$ we impose a cosine-like varying bulk field $H(x)$ which
     introduces a modulation in the concentration of $A$ and $B$ particles. The 
     characteristic length of this perturbation is $\lambda = 32$ lattice spacings. 
     ($a$) Temporal evolution of the concentration of $A$ particles along the $x-$
     direction in the lattice. Times correspond to $t=0$ (circles), $t=10000$ (squares)
     and $t=20000$ (diamonds), respectively. The thick line marks the wavelenght 
     $\lambda = 32$ of the applied bulk field.  
     The temperature is $T=1.5$, and the concentrations are $c_A = c_B = 0.48$.
    The ratio of jump rates is $\Gamma_A / \Gamma_B = 0.01$.
    ($b$) Amplitude of concentration profiles as a function of time, for $A$ particles
    (circles) and $B$ particles (squares). The dashed lines correspond to fits of the
    data to single exponential functions, characterized by a decay constant $D_{int}$.  
          }   \label{fig7}
\end{figure}

Still a different approach was followed to estimate the
interdiffusion constant $D_{int}$. We prepare a system in thermal
equilibrium in the presence of a wavevector-dependent chemical
potential difference $\delta \mu(x)$ defined as
\begin{equation} \label{eq17}
   \delta \mu(x) \equiv \mu_A(x)-\mu_B(x)\equiv \hat{\delta}
   \cos(\frac{2\pi}{\lambda}x),
\end{equation}
$\hat{\delta}$ being an amplitude that needs to be chosen such
that the resulting concentration variation is still in the regime
where linear response holds, and $\lambda$ is the wavelength of
the modulation (which is chosen such that the linear dimension
$L$ of the lattice is an integer multiple of $\lambda$). Note
that in the Ising spin representation $\delta \mu(x)$ simply
translates in a wavelength-dependent magnetic field, of course.
The system then is equilibrated in the presence of this
perturbation for a large number of Monte Carlo steps (of the
order of $10^6$ MCS). This causes a corresponding periodic
concentration variation, see Fig.~\ref{fig7}, left part. The
sinusoidal shape of this initial concentration variation provides
a confirmation that the linear response description is applicable
otherwise the presence of higher harmonics in the concentration
variation would indicate the presence of nonlinear effects. Then
a ``clock'' is set to time $t=0$ and the perturbation $\delta
\mu(x)$ is put to zero for times $t >0$. As a consequence, the
concentration variation decays to zero as the time 
$t \rightarrow \infty$. It turns out that this decay with time can
be described by a superposition of two simple exponential decays,
one governing the decay of the concentration difference $\delta
c(x)=c_A(x)-c_B(x)$ of the particles, the other corresponding to
the decay of the total density. As discussed in detail for the 
random $ABV$ model \cite{17}, the concentration variation can be 
described therefore as ($k=2\pi/\lambda$, and $D_+>D_-$ are two 
diffusion constants)
\begin{equation} \label{eq18}
 \delta c_A(t)= \hat{c}_A^+ \exp(-D_+k^2t)+\hat{c}_A^-\exp(-D_-k^2t)\;,
\end{equation}
\begin{equation} \label{eq19}
 \delta c_B(t)= \hat{c}_B^+ \exp(-D_+k^2t)+\hat{c}_B^-\exp(-D_-k^2t)\;,
\end{equation}
where 
$\hat{c}_A^+, \hat{c}_A^-, \hat{c}_B^+, \hat{c}_B^-$ are amplitude
prefactors, which one can estimate from the treatment that will be
outlined in the following section. Here we only mention that 
$\hat{c}_A^+ + \hat{c}_A^- = \delta c_A(0),\;\hat{c}_B^++\hat{c}_B^-=\delta c_B(0)$,
and in the limit $c_V \rightarrow 0$ we have
$\hat{c}_A^+,\hat{c}_B^+ \propto c_V\rightarrow 0$, while $\hat{c}_A^-,\hat{c}_B^-$ 
stay finite (of the order of $\hat{\delta}$). In this limit the two diffusion 
constants $D_+,D_-$ are of very different order of magnitude, since 
$D_- \propto c_V$, while $D_+$ stays of order unity \cite{17}. Thus
 density variations have a very small amplitude (of order $c_V$)
 and decay fast, while concentration variations decay much slower.
 This consideration leads us to identify $D_-$ as the
 interdiffusion constant $D_{int}$ in this limit. For finite
 nonzero $c_V$, however, in principle both density and
 concentration variations are coupled, and both diffusion
 constants $D_+,D_-$ contribute to the interdiffusion of $A$ and $B$
 particles \cite{17}.

 The right part of Fig.~\ref{fig7} illustrates that even for $c_V$
 as large as $c_V=0.04$ there is already a reasonable separation
 between density and concentration fluctuations: both $\delta
 c_A(t)$ and $\delta c_B(t)$ reach their asymptotic decay (where
 only the same factor $\exp(-D_-k^2t)$ matters, as is evident
 from the fact that there are two parallel straight lines on the
 semilog plot) already at a time $t \approx 2000$, long before the
 concentration variations have decayed to zero.

\section{THEORETICAL PREDICTIONS}
A basic ingredient of all analytical theories are the conservation 
laws for the numbers of A and B particles, which lead to continuity 
equations for the local concentrations $c_A(\vec{r},t),c_B(\vec{r},t)$
\begin{equation} \label{eq20}
  \frac{\partial c_A(\vec{r},t)}{\partial t} + \nabla \cdot
  \vec{j}_A(\vec{r},t)=0,
\end{equation}
\begin{equation} \label{eq21}
  \frac{\partial c_B(\vec{r},t)}{\partial t} + \nabla \cdot
  \vec{j}_B(\vec{r},t)=0,
\end{equation}
Note that these equations hold rigorously, if a local concentration 
field $c_\alpha(\vec{r},t)$ $\{\alpha=A,B\}$ can be defined, unlike 
the so-called constitutive relations, Eqs.~(\ref{eq12}),(\ref{eq13}), 
which are only approximately true: these equations only are supposed 
to hold in the case that the gradients 
$\nabla(\mu_A-\mu_V),\nabla(\mu_B-\mu_V)$ are sufficiently small,
otherwise the relation between currents and gradients is non linear. 
In addition, a second requirement is that statistical fluctuations 
can be neglected; otherwise a random force term needs to be added 
on the right hand side of Eqs.~(\ref{eq12}),(\ref{eq13}) \cite{39}. 
We also note that in our model (unlike real alloys, where vacancies 
can be created by hopping of atoms from lattice sites to interstitial 
sites, and where vacancies can be destroyed by hopping of interstitial 
atoms to a neighboring vacant site of the lattice \cite{1,2,3}) also
vacancies are conserved, and hence
\begin{equation} \label{eq22}
  \frac{\partial c_V(\vec{r},t)}{\partial t} + \nabla \cdot
  \vec{j}_V(\vec{r},t)=0.
\end{equation}
However, as discussed in \cite{17} there is no need to include
$c_V(\vec{r},t)$ and $\vec{j}_V(\vec{r},t)$ as additional
dynamical variables in the problem: the condition that every
lattice site is either occupied by an A-atom, B-atom or vacant
(V) translates into the constraint
$c_A(\vec{r},t)+c_B(\vec{r},t)+c_V(\vec{r},t)=1$. Similarly, one
finds that $\vec{j}_V=-(\vec{j}_A + \vec{j}_B)$ \cite{17}.

In order to be able to relate the chemical potentials in
Eqs.~(\ref{eq12}),(\ref{eq13}) to the concentration variables, we use
the thermodynamic relation
\begin{equation} \label{eq23}
  \mu_\alpha =\left(\frac{\partial F}{\partial N_\alpha}\right)_{{T,N}_
                        {\beta(\neq\alpha)}}\;,
\end{equation}
$N_\alpha$ being the number of particles of species $\alpha$, and $F$ 
being the total free energy of the system. We decompose $F$ into the 
internal energy $U$ and the entropic contribution $-TS$, with $S$ being 
simply the entropy of mixing
\begin{equation} \label{eq24}
    S=-k_B[N_A\ln N_A+N_B \ln N_B+N_V \ln N_V-N \ln N]\;,
\end{equation}
where $N=N_A+N_B+N_V$ then is the total number of sites on the lattice, 
and $c_\alpha =N_\alpha/N$ then is the concentration of species $\alpha$. 
While Eq.~(\ref{eq24}) is exact in the non-interacting $ABV$ model, it
still holds in the disordered phase of the interacting model in the framework
of the Bragg-Williams mean field approximation. In the disordered phase,
no sublattices need to be introduced, and then the concentration variables 
on average are the same for all lattice sites. Then $U$ can be written as
\begin{equation} \label{eq25}
 U = \frac 1 2 Nz(\epsilon_{AA}c_A^2 + 2\epsilon_{AB}c_Ac_B+\epsilon_{BB}c^2_{B}),
\end{equation}
where $z$ is the coordination number of the lattice, and consistent with the 
simulated model (Sec.II) a nearest neighbor interaction is assumed. Note that 
the basic approximation of Eq.~(\ref{eq25}) is the neglect of any correlation 
in the occupancy of neighboring lattice sites.

With some algebra \cite{17} one can reduce Eqs.~(\ref{eq12}), (\ref{eq13}), 
(\ref{eq20})-(\ref{eq25}) to a set of two coupled diffusion equations
\begin{equation} \label{eq26}
  \frac{\partial c _\alpha}{\partial t} = \sum \limits _\beta
  D_{\alpha \beta} \nabla ^2 c_\beta\;,
\end{equation}
where the elements $D_{\alpha \beta}$ of the diffusion matrix are given by
\begin{equation} \label{eq27}
 D_{AA}=\Lambda_{AA}(\frac{1}{c_A}+\frac{1}{c_V}+\frac{z\epsilon_{AA}}{k_BT}) 
      + \Lambda _{AB}(\frac{1}{c_V} + \frac{z\epsilon_{AB}}{k_BT}) \;,
\end{equation}
\begin{equation} \label{eq28}
 D_{AB}=\Lambda_{AA}(\frac{1}{c_V} + \frac{z\epsilon_{AB}}{k_BT}) + 
   \Lambda_{AB}(\frac{1}{c_B} + \frac{1}{c_V} + \frac {z\epsilon_{AA}}{k_BT}) \;,
\end{equation}
\begin{equation} \label{eq29}
   D_{BA} = \Lambda_{AB}(\frac{1}{c_A}+\frac{1}{c_V}+\frac{z\epsilon_{AA}}{k_BT}) 
          + \Lambda_{BB}(\frac{1}{c_V}+\frac {z\epsilon_{AB}}{k_BT}) \;,
\end{equation}
\begin{equation} \label{eq30}
  D_{BB}=\Lambda_{AB}(\frac{1}{c_V}+\frac{z\epsilon_{AB}}{k_BT})
        +\Lambda_{BB}(\frac{1}{c_B}+\frac{1}{c_V} 
                                   + \frac{z\epsilon_{BB}}{k_BT}) \;.
\end{equation}
Note that $D_{AB} \neq D_{BA}$. Introducing Fourier transforms and
diagonalizing the diffusion matrix the solution indeed can be cast into 
the form of Eqs.~(\ref{eq18}),(\ref{eq19}). As has already been mentioned 
in this context, for $c_V\rightarrow  0 $ the two eigenvalues $D_+,D_-$ of 
the diffusion matrix adopt very different orders of magnitude \cite{17}:
\begin{equation} \label{eq31}
   D_+ \approx (\Lambda_{AA}+2\Lambda_{AB}+ \Lambda_{BB})/c_V,
\end{equation}
\begin{equation} \label{eq32}
  D_- \approx  \frac{\Lambda_{AA}\Lambda_{BB}-\Lambda^2_{AB}}
       {\Lambda_{AA}+2\Lambda_{AB}+\Lambda_{BB}}
       (\frac{1}{c_A} + \frac{1}{c_B} - \frac{2z\epsilon}{k_BT}).
\end{equation}
Since in this limit the $\Lambda_{\alpha \beta} \propto c_V$, the coefficient 
$D_+$ reaches a finite limit for $c_V \rightarrow 0$, while $D_-\propto c_V$. 
We also recognize that $D_-$ can be decomposed into a product of two factors: 
a ``kinetic factor'' $\Lambda_{int}$, composed by a combination of Onsager 
coefficients, and a ``thermodynamic factor'', which is nothing but an effective 
inverse ``susceptibility'' $\chi^{-1}$ describing concentration fluctuations, 
normalized per lattice site,
\begin{equation} \label{eq33}
  \chi^{-1}=c_A^{-1}+c_B^{-1}-2z\epsilon/k_BT=[c_A(1-c_A)]^{-1}-2z\epsilon/k_BT
\end{equation}
In the last step, we used the fact that $c_B=1-c_A$ for $c_V\rightarrow 0$. 
We call $\chi$ a ``susceptibility'' because in the translation to the Ising 
spin representation $\chi$ simply becomes proportional to the derivative of 
the ``magnetization'' with respect to the field. Note that for $\epsilon > 0$ 
(i.e., a mixture with unmixing tendency) Eq.~(\ref{eq33}) exhibits a vanishing 
of $\chi^{-1}$ and hence of the interdiffusion constant $D_-$ at the mean field 
spinodal curve, defined by
\begin{equation} \label{eq34}
   k_BT_s(c_A)/\epsilon = 2c_A(1-c_A)z \;.
\end{equation}
The mean field spinodal touches the coexistence curve of such a phase-separating 
mixture at its maximum in the critical temperature, i.e. 
$k_BT_c^{MF}/\epsilon=z/2=2$, for a square lattice. Actually, the symmetry of the 
Ising Hamiltonian in zero field implies that the maximum critical temperature of 
the Ising antiferromagnet, which occurs at zero field as well, then is also
given by
\begin{equation} \label{eq35}
    k_B \, T_{\,c,max}^{\,MF} /|\epsilon|=z/2=2, \quad \epsilon <0
\end{equation}
Comparing this estimate to the exact result, Eq.(~\ref{eq2}), we notice that 
the mean field approximation actually overestimates the maximum critical 
temperature of the ordering alloy by almost a factor of two, as is well known. 
Note that this error increases for $c_A \neq 1/2$ \cite{35}. So Eq.~(\ref{eq32})
cannot be assumed to be quantitatively reliable. Note that for ordering alloys 
(where $\epsilon <0$) the interdiffusion constants gets enhanced (rather than 
reduced, as it happens for alloys with unmixing tendency) as an effect of the 
interactions. Beside that, Eq.~(\ref{eq32}) does not predict any singularity 
of $D_-$ as one approaches the order-disorder phase boundary $T_c(c_A)$ from the
disordered side.

Discussing now the kinetic factor $\Lambda_{int}$, we recall the popular 
approximation to neglect the off-diagonal Onsager coefficient in comparison 
to the diagonal ones. This leads to
\begin{equation} \label{eq36}
  \Lambda_{int} \approx \Lambda_{AA}^{-1} + \Lambda_{BB}^{-1}
\end{equation}
With this approximation, Eq.~(\ref{eq32}) reduces to the well-known 
``slow mode theory '' of interdiffusion, which has been much debated in the case 
of fluid polymer mixtures \cite{40}-\cite{45}. A mean field type approximation 
for self-diffusion \cite{17,40,41,42,43} then relates the Onsager coefficients
$\Lambda_{AA},\Lambda_{BB}$ and tracer diffusion coefficients $D_t^A,D_t^B$,
\begin{equation} \label{eq37}
   \Lambda_{AA}=D_t^A c_A\;,\quad \Lambda_{BB}=D_t^Bc_B
\end{equation}
and thus the ``slow mode'' theory predicts the following relation
between tracer diffusion coefficients and the interdiffusion
constant (remember $c_B=1-c_A$ for $c_V \rightarrow 0$)
\begin{equation}\label{eq38}
D_{int}^{s.m.}=\{(D_t^A
c_A)^{-1}+[D_t^B(1-c_A)]^{-1}\}\{[c_A(1-c_A)]^{-1}-2z\epsilon/k_BT\}.
\end{equation}
A rather different result, the so-called ``fast mode'' theory \cite{44,45}, 
can be obtained by several distinct arguments. We mention only one of these 
arguments here, which starts from the assumption \cite{44} that everywhere 
the vacancy concentration $c_V(\vec{r},t)$ is in thermal equilibrium, i.e.
\begin{equation} \label{eq39}
  \nabla \mu_v=0.
\end{equation}
Of course, in our model Eq.~(\ref{eq39}) cannot be justified, in view 
of the constraints $c_V(\vec{r},t)=1-c_A(\vec{r},t)-c_B(\vec{r},t)$,
$\vec{j}_V=-(\vec{j}_A+\vec{j}_B)$ and Eqs.~(\ref{eq22})-(\ref{eq24}) 
there is no freedom to make additional assumptions on $\mu_V$ at all, 
$\nabla\mu_V(\vec{r},t)$ already is determined from these other equations. 
However, the motivation for Eq.~(\ref{eq39}) is that for real systems 
there is no strict conservation for the number of vacancies: in real 
(three-dimensional) alloys, vacancies can be created and destroyed by 
formation or annihilation of interstitial atoms, or by interaction with 
other lattice imperfections such as dislocations, grain boundaries, etc. 
For two-dimensional surface alloys \cite{20}, vacancies can be created 
and destroyed if an atom from the considered surface monolayer becomes 
an adatom on top of this monolayer, or an adatom executing surface diffusion
\cite{18,30} becomes incorporated into the monolayer via a jump to a vacant 
site inside the monolayer. In view of these physical mechanisms which are 
forbidden in our model, Eq.~(\ref{eq39}) may represent a physically interesting 
limiting case. A priori, it is not clear for a particular system, whether 
for the time scales of interest it is closer to a situation where vacancies are 
in equilibirum (Eq.~(\ref{eq39})) or conserved (Eq.~(\ref{eq22})). Our numerical 
studies are concerned with the latter case exclusively. Nevertheless it is of 
interest to mention that Eq.~(\ref{eq39}) yields also a structure 
$D_-=\Lambda_{int}\chi^{-1}$ but with $\Lambda_{int}=c_BD_t^A +c_AD_t^B$ and 
hence one finds instead of Eq.~(\ref{eq38}) \cite{17}
\begin{equation} \label{eq40}
  D_{int}^{f.m.}=[(1-c_A)D_t^A+c_AD_t^B]\{[c_A(1-c_A)]^{-1}-2z\epsilon/k_BT\}.
\end{equation}
While for $D_t^B \gg D_t^A$ (a case expected if $\Gamma_A \ll \Gamma_B$, as used 
in our simulation) one expects from Eq.~(\ref{eq40}) that the faster diffusing $B$ 
species dominates interdiffusion, the opposite is true according to 
Eq.~(\ref{eq38}): therefore the names ``fast mode'' and ``slow mode'' theory have
been chosen. In both equations (and in Eq.~(\ref{eq32}), where the off-diagonal 
Onsager coefficient is not neglected, unlike in both these theories) the 
thermodynamic factor is treated by a simple Bragg-Williams mean field approximation, 
however, which is no problem for the random alloy $ABV$ problem treated in
Ref.~\cite{17}, but clearly will introduce additional shortcomings in the present 
case.

As a final disclaimer of this section we emphasize that 
Eqs.~(\ref{eq20})-(\ref{eq40}) were meant to provide a brief review of 
``chemical diffusion'' (or ``collective diffusion'') in the context of 
the present lattice gas model only, and hence many interesting and 
important facets of this topic have not been mentioned at all and we 
direct the interested reader to the rich literature on this subject 
\cite{1,2,3,4,5,6,7,8,9,46,47}.

\section{SIMULATION RESULTS}
\subsection{Tracer diffusion}
\begin{figure}
  \begin{center}
     \scalebox{0.55}{\includegraphics{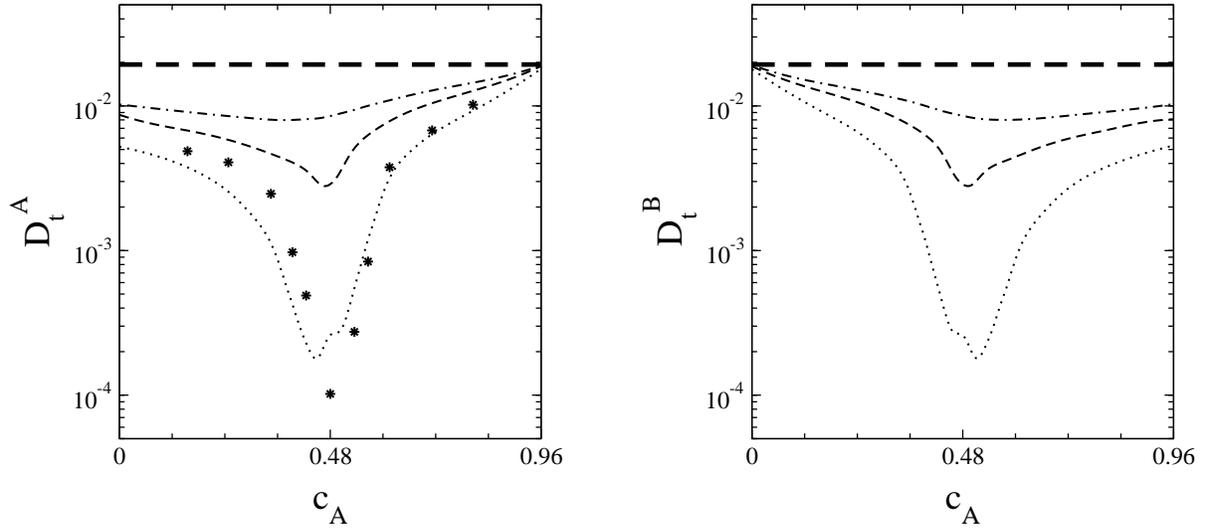}}
  \end{center}
  \caption{
        Tracer diffusion coefficients for $A$ particles (left) and $B$ particles 
        (right), as a function of the concentration $c_A$. The jump rates are
        $\Gamma_A = \Gamma_B = 1$ and several temperatures are considered: $T=0.6$ 
        (dotted line), $T=0.912$ (dashed line) and $T=1.2$ (dot-dashed line). 
         The thick line indicates in both cases the noninteracting,
         infinite temperature limit (\emph{random alloy model}).  
         Dots represent results obtained from Eq.~(\ref{eq41}).
          }   \label{fig8}
\end{figure}

\begin{figure}
  \begin{center}
     \scalebox{0.55}{\includegraphics{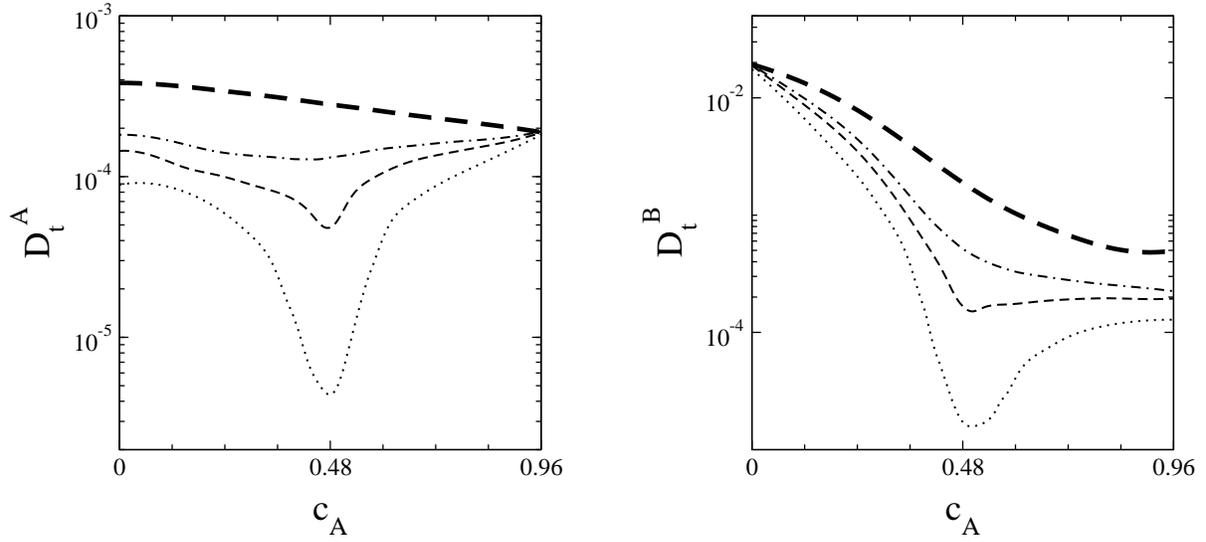}}
  \end{center}
  \caption{
      Tracer diffusion coefficients for $A$ particles (left) and $B$ particles 
      (right), as a function of the concentration $c_A$. The jump rates are
       $\Gamma_A / \Gamma_B = 0.01$ and several temperatures are considered: $T=0.6$ 
      (dotted line), $T=0.912$ (dashed line) and $T=1.2$ (dot-dashed line). 
       The thick line indicates in both cases the noninteracting,
       infinite temperature limit (\emph{random alloy model}). 
        }   \label{fig9}
\end{figure}

We start with a discussion of the tracer diffusion coefficients 
(Figs.~\ref{fig8}, \ref{fig9}). The simplest case refers to equal jump rates 
$\Gamma_A=\Gamma_B=1$ of both types of particles $A$ and $B$ (Fig.~\ref{fig8}). 
In the infinite temperature limit then there is no longer any physical 
difference between $A$ and $B$ particles, they simply differ only by their labels: 
then $D_t^A=D_t^B$, and become independent of concentration $c_A$ (thick 
horizontal straight line in Fig.~\ref{fig8}). Note that for $c_A=0.96$ there are 
no $B$ particles since $c_V=0.04$ and then $D_t^A$ becomes independent of 
temperature, similarly as $D_t^B$ becomes independent of temperature for $c_A=0$. 
Of course, the curves for $D_t^B$ are simply the mirror images of those for 
$D_t^A$ around the symmetry line $c_{A,max}^{crit}=(1-c_V)/2=0.48$ of the
static phase diagram, Fig.~\ref{fig2}b, since an interchange of $A$ and $B$ means
that $c_A$ gets replaced by $1-c_V-c_B$.

It is seen that the onset of ordering depresses self-diffusion very strongly, 
while short range order (as it occurs for $T=1.2$) has a minor effect only. 
For $T=0.6$, however,the ordering near $c_A=0.48$ is rather perfect and there 
deep minima of $D_t^A,D_t^B$ occur, the tracer diffusion coefficients decrease
by about two orders of magnitude. Of course, since $D_t^A,D_t^B$ are 
\textit{not symmetric} around $c_{A,max}^{crit}=c_{B,max}^{crit}=(1-c_V)/2=0.48$, 
due to the choice of a kinetic Monte Carlo algorithm which lacks the symmetry 
between the motion of an $A$ particle, mediated by a vacancy, in a $B$ environment 
and in an $A$ environment at finite temperatures, the minimum of $D_t^A$ does
not occur precisely at $c_{A,max}^{crit}$, as is seen from Fig.~\ref{fig8}
(left part). In our algorithm, an $A$ particle jumps to a vacant site with a
jump rate $\Gamma_A \exp(-\Delta n|\epsilon_{AB}|/k_BT)$ when the difference 
between the number of $AB$ bonds involving an energy $\epsilon_{AB}$ each
between the initial and final state is $\Delta n>0$, and with a jump rate 
$\Gamma _A$ else.
It is easy to be convinced that this algorithm satisfies detailed balance with 
the canonic equilibrium distribution, as it should be \cite{19}. In the limit 
$c_A\rightarrow 1-c_V$, we always have $\Delta n=0$, so there is no temperature 
dependence. In the limit $c_A\rightarrow 0$, however, every $A$ atom not having 
a vacancy as nearest neighbors will have four $B$ neighbors on the square lattice, 
while an $A$ atom with a vacancy neighbor has only three $B$ neighbors. As a result, 
the jump of an $A$ atom that has a $B$ neighbor, to a vacant site involves ``breaking'' 
an $AB$ bond, and hence this rate is suppressed by a factor 
$\exp(-|\epsilon_{AB}|/k_BT)$. This effect is responsible for the temperature 
dependence of $D_t^A$ for $c_A\rightarrow 0$.

Kehr et al. \cite{17} presented arguments to relate the tracer diffusion 
coefficients to Onsager coefficients which take a simple form in the case of 
identical jump rates $\Gamma_A,\Gamma_B$, namely
\begin{equation} \label{eq41}
  D_t^A=\Lambda_{AA}/c_A - \Lambda_{BA}/c_B, \quad
  D_t^B=\Lambda_{BB}/c_B-\Lambda_{AB}/c_A
\end{equation}
Using our estimates for the Onsager coefficients at $T=0.6$ (see below) in 
Eq.~(\ref{eq41}), one sees that the trend of the concentration dependence of 
$D_t^A$ is reproduced rather well. However, one should note that the derivation 
of Eq.~(\ref{eq41}) is rigorous only for the special case $\epsilon_{AB}/k_BT=0$, 
because only then the distinction between $A$ and $B$ particles forming the 
environment of a tagged $A$ particle can be neglected.

When $\Gamma _A \neq \Gamma_B$ the self-diffusion coefficients $D_t^A$ and $D_t^B$ 
lack any symmetric relation of their concentration dependence already in the 
random alloy limit \cite{17}, and for $\epsilon_{AB}/k_BT < 0$ we are not aware of 
any theoretical treatment to which our simulation results (Fig.~\ref{fig9}) could
be compared. 
Interestingly, for not too low temperatures (such as $T=0.912$, $T=1.2$) the 
concentration dependence of $D_t^A$ (the slower diffusing species, since 
$\Gamma_A/\Gamma_B=0.01$ has been chosen in Fig.~\ref{fig9}) is rather weak 
throughout, while for $D_t^B$ we have a strong decrease when $c_A$ increases up to 
about $c_{A,max}^{crit} = 0.48$. For $c_A > c_{A,max}^{crit}$ again a very
weak concentration dependence results. 
For $T=0.6$ again pronounced minima near $c_{A,max}^{crit}$ are seen. Now, for 
$D_t^B$ we have a strong decrease when $c_A$ increases up to about
$c_{A,max}^{crit}$, while for $c_A > c_{A,max}^{crit}$ again a very weak
concentration dependence results. 

\begin{figure}
  \begin{center}
     \scalebox{0.55}{\includegraphics{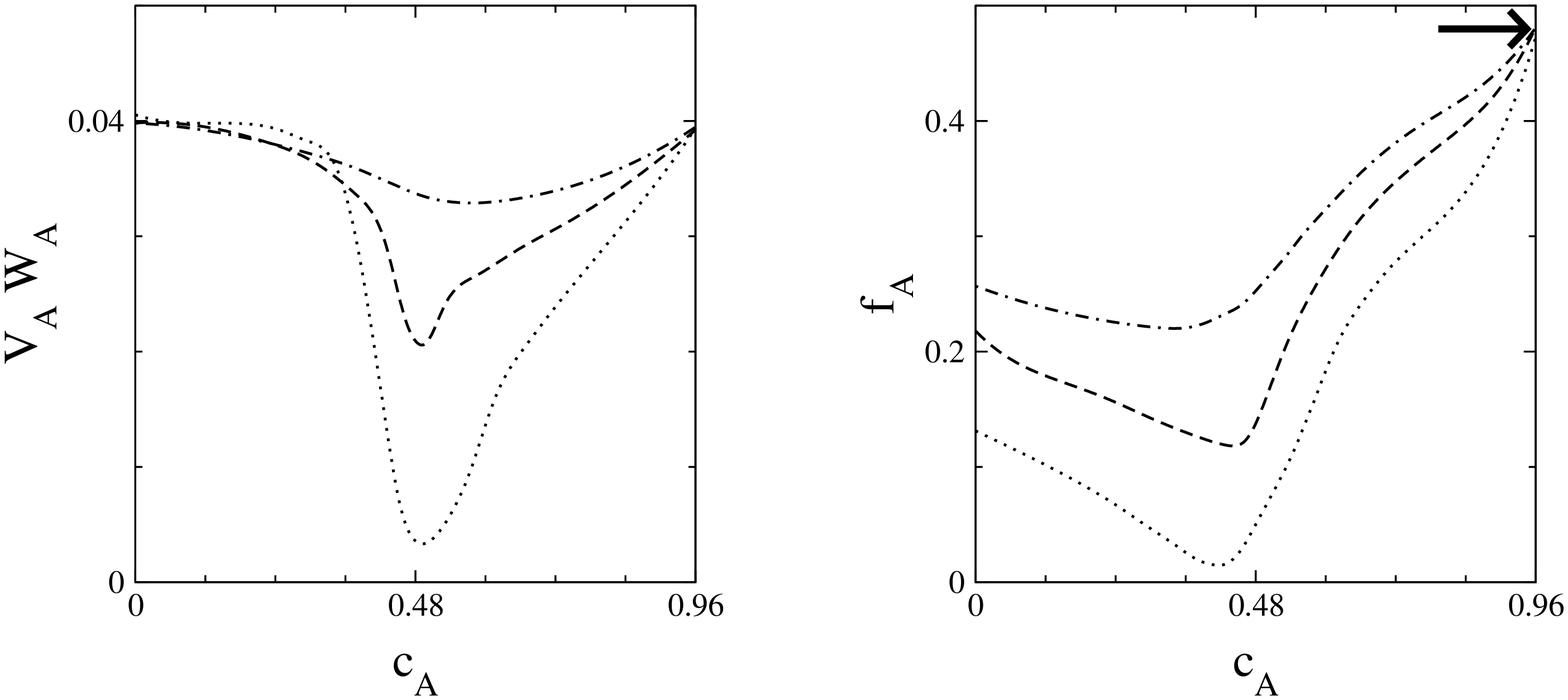}}
  \end{center}
  \caption{
      Effective jump rate (left) and correlation factor (right) for $A$ particles,
      as a function of the concentration $c_A$ and for different temperatures:
       $T=0.6$ (dot line), $T=0.91$ (dashed line) and $T=1.2$ (dot-dashed line). 
      The jump rates are $\Gamma_A = \Gamma_B = 1$. \\ 
      The first quantity provides an idea of the rate
      at which jumps \emph{actually} occur at a certain temperature and composition.
      It is defined as the product of the vacancy availability factor $V_A$ (related to
       the short-range order parameter $\alpha_1$) and the average jump rate $W_A$ 
       (defined as the quotient of the number of performed jumps to the number of
        all attempted jumps). The $T \rightarrow \infty$ limit is given by 
          $V = c_v \equiv 0.04$, because in this case $\alpha_1 \equiv 0$. \\
       Once we obtain $V W$ we can estimate the correlation factor $f$ applying
       the definition $D_t = V W f$ and using $D_t$ from Fig~\ref{fig8}. 
       See Refs. \cite{23,24,25,26} for details on the effect of correlations
       on tracer diffusion in lattice gas models. \\
        The limit value $f=0.487$ for $c_A \rightarrow 0.96$ is known from Ref. 
        \cite{28}. This corresponds to a noninteracting, one-component 
        lattice gas in a square lattice with concentration $c=0.96$.
          }   \label{fig10}
\end{figure}

Moreover, when for $c_A = c_{A,max}^{crit}$ the order of the $AB-$ checkerboard 
structure is perfect (apart from a four per cent of vacant sites in the system), 
a jump of an atom to a vacant site occurs with rates 
$\Gamma_A \exp(3\epsilon_{AB}/k_BT)$ or $\Gamma_B \exp(3\epsilon_{AB}/k_BT)$, 
respectively, while the backward jump occurs at rates $\Gamma_A,\Gamma_B$. As
a result, a high probability for backward jumps is expected, and this is borne 
out by a study of the correlation factor $f$ for self-diffusion (Fig.~\ref{fig10}, 
right part). Following standard treatments \cite{1,2,3,4,5,6,23} we decompose 
tracer diffusion coefficients $D_t$ as
\begin{equation} \label{eq42}
   D_t=VWf\;,
\end{equation}
where $V$ is the vacancy availability factor already defined in Eq.~(\ref{eq9}), and 
$W$ is the average jump rate for the considered particle species. $W$ is easily 
estimated in the simulation from the ratio of the number of performed jumps to the 
number of all attempted jumps. The product $V_AW_A$ is plotted in Fig.~\ref{fig10} 
(left part) versus $c_A$ at various temperatures. For 
$\epsilon_{AB}/k_BT \rightarrow 0$ we simply expect a horizontal straight line, 
$V_A W_A = 0.04$, since then $W_A=1$, $V_A = c_V$ ($\alpha_1=0$). There is no 
independent way to determine $f$, however. Therefore Eq.~(\ref{eq42}) is taken as 
a definition of $f$, to be derived from $D_t$, while the tracer diffusion constants 
are estimated from the mean square displacements of the tagged particles, as explained 
in Sec.~\ref{secIII} of this paper. For $c_A \rightarrow 0.96$, when no $B$ particles 
are present, the temperature dependence drops out and $f$ reduces to the value 
$f=0.487$ known from studies of a one-component non-interacting lattice gas on
a square lattice with concentration $c_A=0.96$ \cite{28}. Note that our data for 
$D_t$, $V$, $W$ and $f$ at the higher temperatures (where no order-disorder transition 
occurs) resemble analogous results of Murch \cite{25} for a simple cubic alloy.

\begin{figure}
  \begin{center}
     \scalebox{0.55}{\includegraphics{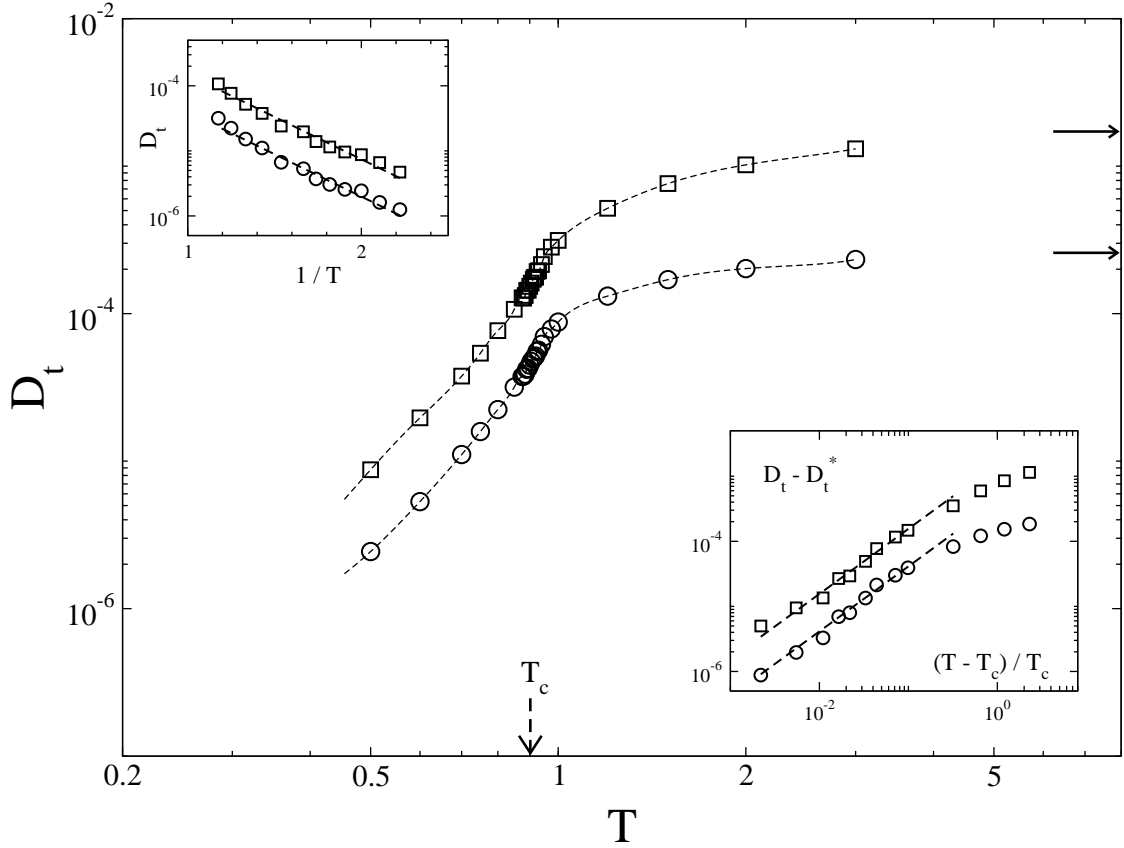}}
  \end{center}
  \caption{
       Dependence of the tracer diffusion coefficients on the temperature,
   for a stoichiometric composition $c_A = c_B = 0.48$. The ratio of jump rates
     is $\Gamma_A / \Gamma_B = 0.01$. Circles are $A$ particles and squares are
     $B$ particles. \\
     The dashed arrow marks the critical temperature $T_c = 0.905$ (in units of 
      the Ising critical temperature), while the thick arrows indicate the 
       asymptotic, infinite temperature values for both coefficients. \\
     Inset (up): Arrhenius plot of $D_t$ for $T \ll T_c$.
     Inset (bottom): scaling plot of $|D_t - D_t^*| \sim |T - T_c|^{1-\alpha}$ with
     $\alpha = 0$ (specific heat exponent of the Ising model). The dashed line
      has a slope of unity. 
       }   \label{fig11}
\end{figure}

As a final comment about self-diffusion, we consider the temperature dependence of 
$D_t^A$ and $D_t^B$ for the critical concentration $c_A^{crit}=c_B^{crit}=0.48$ 
(Fig.~\ref{fig11}). One sees that at high temperatures ($T \geq 2T_c$) the temperature
dependence is very weak, and the tracer diffusion coefficients settle down at their 
infinite temperature asymptotes. Approaching the critical point one sees a more rapid 
decrease of both $D_t^A$ and $D_t^B$, with a maximum slope presumably right at $T_c$, 
while for $T$ below $T_c$ a crossover to the expected thermally activated behavior at 
low temperatures occurs. In fact, one expects that 
$D_t-D_t^* \propto (T-T_c)^{1-\alpha}$ \cite{24}, where $D_t^*$ is the value of the 
tracer diffusion coefficient at the critical point, and $\alpha$ is the specific heat 
exponent of the model. However, for the two-dimensional Ising model $\alpha =0$ 
\cite{32,33,34}, i.e. the specific heat has a logarithmic singularity only. The insert 
of Fig.~\ref{fig11} shows a log-log plot of $D_t-D_t^*$ versus $(T-T_c)/T_c$, and one 
sees that the data are compatible with a power law with slope of unity; presumably the 
accuracy of our simulations does not suffice to identify the presence of a
logarithmic singularity in our data.

\subsection{Onsager coefficients}
\begin{figure}
  \begin{center}
     \scalebox{0.6}{\includegraphics{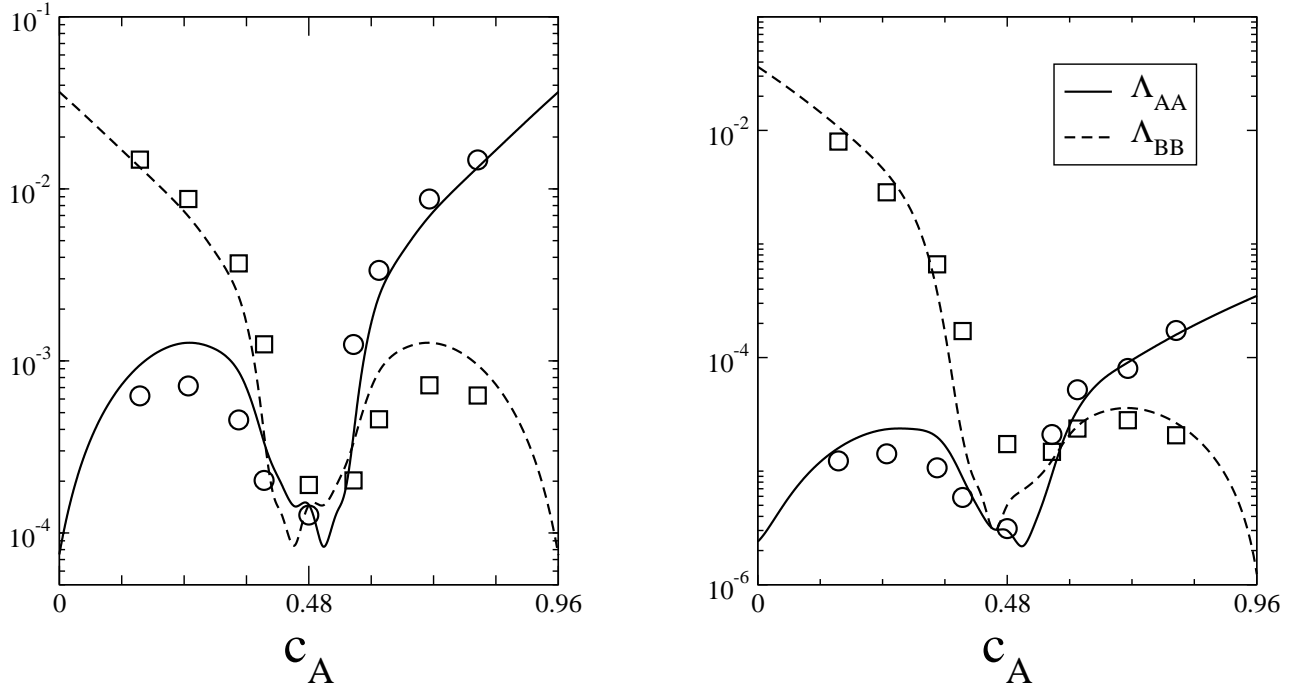}}
  \end{center}
  \caption{
     Plot of the Onsager coefficients $\Lambda_{ii}$ as a function of the
    concentration $c_A$, for a fixed temperature $T=0.6$. The jump rates are
     $\Gamma_A = \Gamma_B = 1$ (left) and $\Gamma_A / \Gamma_B = 0.01$ (right). 
     The lines correspond to data obtained directly
    from the simulations, while the points correspond to the estimates obtained
    after using the aproximation of Eq.~(\ref{eq43}) for the random alloy model. 
       }   \label{fig12}
\end{figure}

\begin{figure}
  \begin{center}
     \scalebox{0.6}{\includegraphics{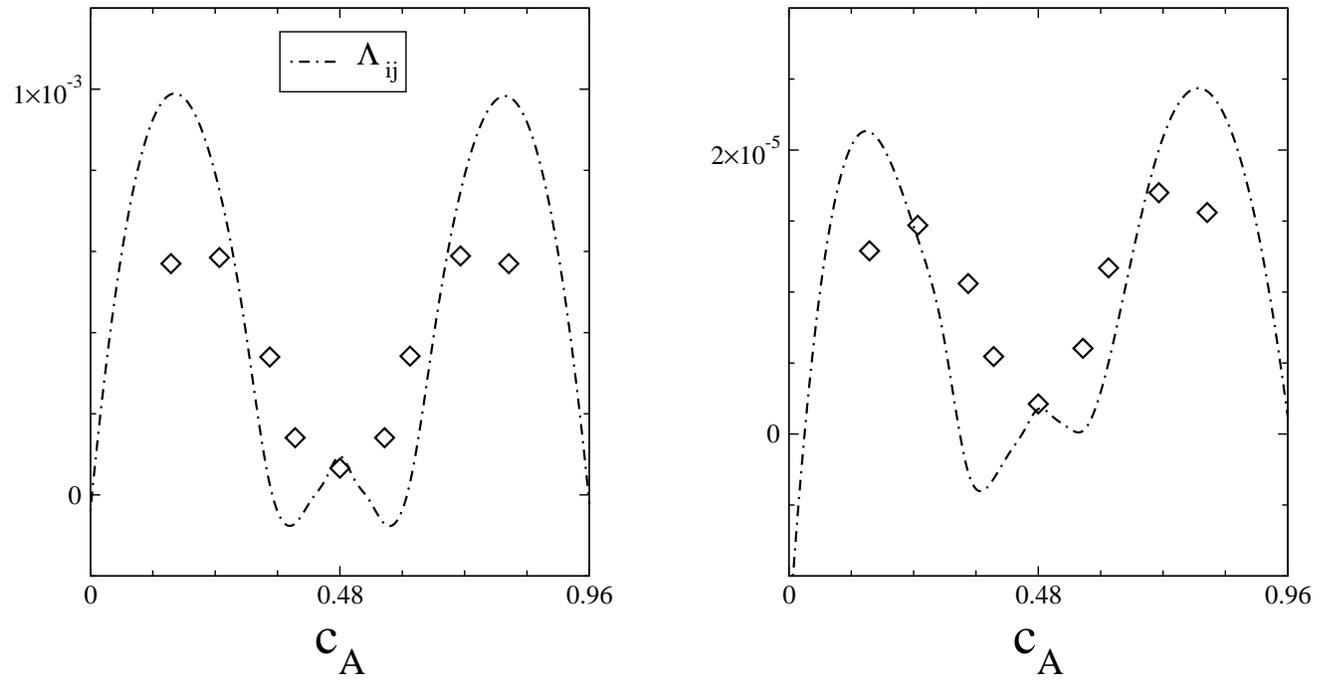}}
  \end{center}
  \caption{
       The same as figure \ref{fig12} but here the crossed coefficient 
       $\Lambda_{ij}$  is shown.
       }   \label{fig13}
\end{figure}

As a first issue of this subsection, we turn to the concentration dependence of the 
Onsager coefficients (Figs.~\ref{fig12}, \ref{fig13}). For $\Gamma_A = \Gamma_B$ all
Onsager coefficients are symmetric around $c_A=c_B=(1-c_V)/2$, as it must be, while 
for $\Gamma _A \neq \Gamma _B$ they are not. We have also included an approximate 
relation suggested by Kehr et al. \cite{17} between Onsager coefficients and tracer 
diffusion coefficients, namely
\begin{equation} \label{eq43}
   \Lambda_{\alpha \beta} = c_\alpha D_t^\alpha \;\left[\delta _{\alpha
   \beta}+ \frac {1-f(c)}{f(c)} \, \frac{c_\beta D_t^\beta}{\sum
   \limits_\gamma c_\gamma D_t^\gamma}\right],
\end{equation}
where $c=c_A+c_B$, $f(c)$ being the correlation factor for tagged-particle diffusion 
in a lattice gas with summary concentration $c$. It is seen that this relation 
accounts for the general trend of the diagonal Onsager coefficients rather well,
although for the off-diagonal Onsager coefficient it seems to work only qualitatively 
(Fig.~\ref{fig13}). In the regime of the ordered phase the diagonal Onsager 
coefficients (note the logarithmic ordinate scale) are distinctly smaller than for 
$c_A\rightarrow 0$ or $c_B \rightarrow 0$, respectively, when $\Gamma_A = \Gamma_B$.

An interesting aspect of the off-diagonal Onsager coefficient
$\Lambda_{AB} = \Lambda_{BA}$ (Fig.~\ref{fig13}) is that it is essentially zero for 
$c_A \rightarrow 0$ if $\Gamma_A=\Gamma_B$ while for $\Gamma_A/\Gamma_B=0.01$ it is 
essentially negative in this limit. A negative Onsager coefficient means that the 
currents of $A$ and $B$ particles are oriented in the opposite direction. A further 
change of sign of this off-diagonal coefficient is found near the phase boundary of the 
order-disorder transition; but near $c_A=c_B=(1-c_V)/2$ the Onsager coefficient seems 
to be positive again, although its absolute value seems to be very small. We do not 
have any clear physical interpretation for this surprising behavior. Note also, that 
Eq.~(\ref{eq43}) can never yield a negative Onsager coefficient, since 
$0 \leq f(c)\leq 1$ by definition, and hence all terms in Eq.~\ref{eq43} are
non-negative.

Finally Fig.~\ref{fig14} shows the temperature dependence of the Onsager coefficients 
for the concentration $c_A=c_B=(1-c_V)/2$ where the critical temperature $T_c$ of the 
order-disorder transition is maximal. Note that for $\Gamma_A/\Gamma_B =0.01$ the
magnitude of the off-diagonal Onsager coefficient $\Lambda_{AB}$ is comparable  
to the smaller ($\Lambda_{AA}$) of the diagonal ones, both at very high and at very 
low temperatures. This finding confirms the conclusion of Kehr et al. \cite{17},
that in general the off-diagonal Onsager coefficient must not be neglected. We also 
note that the general trend of the temperature dependence of the Onsager coefficients 
is very similar to the behavior of the self-diffusion coefficient, see 
Fig.~\ref{fig11}. Both quantities reflect the strong decrease of mobility of the
particles at low temperatures.

\subsection{Interdiffusion}
Fig~\ref{fig15} presents a plot of the interdiffusion constant $D_{int}$ vs. 
concentration for the case of equal jump rates ($\Gamma_A=\Gamma_B=\Gamma =1$) 
at $T=0.6$ and compares the results to various analytical approximations:
$D_-$ (Eq.~(\ref{eq32})), the ``slow mode'' expression $D_{int}^{s.m.}$
(Eq.~(\ref{eq38})), the ``fast mode'' expression $D_{int}^{f.m.}$ (Eq.~(\ref{eq40})), 
and a very simple result justified by Kehr et al. \cite{17} for the non-interacting 
random alloy model,
\begin{equation} \label{eq44}
    D_{int}^{n.i.}=(1-c)\,f(c)\, \Gamma\;.
\end{equation}
While this last expression overestimates the numerical results, all other expressions 
underestimate them significantly. It is seen that in this case there is not much 
difference between the slow mode and fast mode theory, but both are off from the data. 
In this case using the full expression (Eq.~(\ref{eq32})) presents no improvement, 
unlike the non-interacting case. Of course, at finite temperature in $d=2$ the mean 
field theory implicit in Eq.~(\ref{eq32}) is not expected to be accurate at all.

It now is no surprise any longer that in the asymmetric case $\Gamma_A/\Gamma_B=0.01$ 
the various approximate expressions are not reliable either (Fig.~\ref{fig16}). In 
particular, for concentrations near $c_A=c_B=(1-c_V)/2=0.48$ a pronounced minimum is 
predicted, while the actual simulation results reveal a rather shallow minimum only. 
Again the conclusion is that there is no reliable simple relation between self-diffusion
and interdiffusion coefficients, and the temperature dependence of $D_{int}$ at 
$c_A=c_B=0.48$ at higher temperatures (Figs.~\ref{fig17},\ref{fig18}) confirms this 
conclusion. Again, for $\Gamma_A=\Gamma_B=\Gamma=1$  Eq.~(\ref{eq44}) is closest to the
data, while Eq.~(\ref{eq32}) is worst. For $T \rightarrow \infty$, however, in this 
limit for $c_A=c_B=(1-c_V)/2$ and $\Gamma_A=\Gamma_B=1$ all expressions coincide 
(at a value highlighted by an arrow in Fig.~\ref{fig17}), and the numerical data have 
been found in good agreement with this prediction \cite{17}. Thus it is clear that 
including interactions among the particles destroys the applicability of the simple 
theories.

\section{CONCLUSIONS}
In this paper, the study of mobility of particles, interdiffusion and tracer 
diffusion coefficients of a lattice model for a binary alloy, that was presented 
in Ref.~\cite{17} for the simple non-interacting limit only, has been extended 
to the case where an attractive nearest-neighbor interaction between unlike particles
leads to an order-disorder transition on the considered square lattice. While most 
theoretical considerations of the previous work \cite{17} can be simply extended to 
the present case, the mean-field character of the approximations that are involved
clearly emerges as a severe limitation of the usefulness of all these approaches. 
On the other hand the Monte Carlo techniques described in Ref.~\cite{17}, suitable 
for the direct estimation of all Onsager coefficients and the interdiffusion constant
$D_{int}$ as function of the ratio of jump rates $\Gamma_A/\Gamma_B$, temperature $T$ 
and concentration $c_A$, are rather straightforward to apply.
Exploring this rather large parameter space numerically is however somewhat tedious, 
and an understanding of diffusion phenomena within the framework of lattice models for 
interacting particles by simple analytical expressions clearly would be desirable. 
However, the approximate expressions discussed in the present paper clearly do not give
qualitatively accurate results.

Of course, the present study is a first step only: in order to make closer contact
with possible experiments in surface layers of metallic alloys, it would be interesting
to consider other lattice symmetries (triangular and centered rectangular lattice rather
than square lattices), further neighbors interactions, etc..

A very important extension would also be the inclusion of asymmetric effects 
($\epsilon_{AA} \neq \epsilon_{BB}$) and nonzero energy parameters involving
vacancies ($\epsilon_{AV}, \epsilon_{BV}, \epsilon_{VV}$). Thus effects could
be described that vacancies occupy preferentially sites at interfaces \cite{21}
or in one of the sublattices \cite{ecm}. Such effects are expected to modify the
diffusion behavior significantly.

We thus hope the present study will stimulate the development of more accurate 
theoretical descriptions of diffusion phenomena in alloys that undergo order-disorder 
transitions. Also corresponding experiments studying a wide range of temperature and 
composition, would be desirable. Then it might be worthwhile to combine the present 
kinetic Monte Carlo methodology with ``ab initio'' calculation of jump rates, ordering 
energies $\epsilon_{\alpha\beta}$, etc.

\underline{Acknowledgements}: 
One of us (A.D.V.) is grateful to the German Academic Exchange Service (DAAD) 
and to the Deutsche Forschungsgemeinschaft (DFG), grant No SFB TR6/A5, for 
financial support.

%
%
\newpage

\begin{figure}
  \begin{center}
     \scalebox{0.55}{\includegraphics{onsa_ca0.48_g0.01a.eps}}
  \end{center}
  \caption{
      Plot of the Onsager coefficients $\Lambda_{ij}$ as a function of the 
     temperature, for a stoichiometric concentration $c_A = c_B = 0.48$ and
     jump rates $\Gamma_A / \Gamma_B = 0.01$. The lines are drawn to guide the
     eyes. \\
     Inset: plot of the same coefficients in the vicinity of the critical temperature,
     showing the linear approach to finite values right at $T_c$.
       }   \label{fig14}
\end{figure}

\begin{figure}
  \begin{center}
     \scalebox{0.55}{\includegraphics{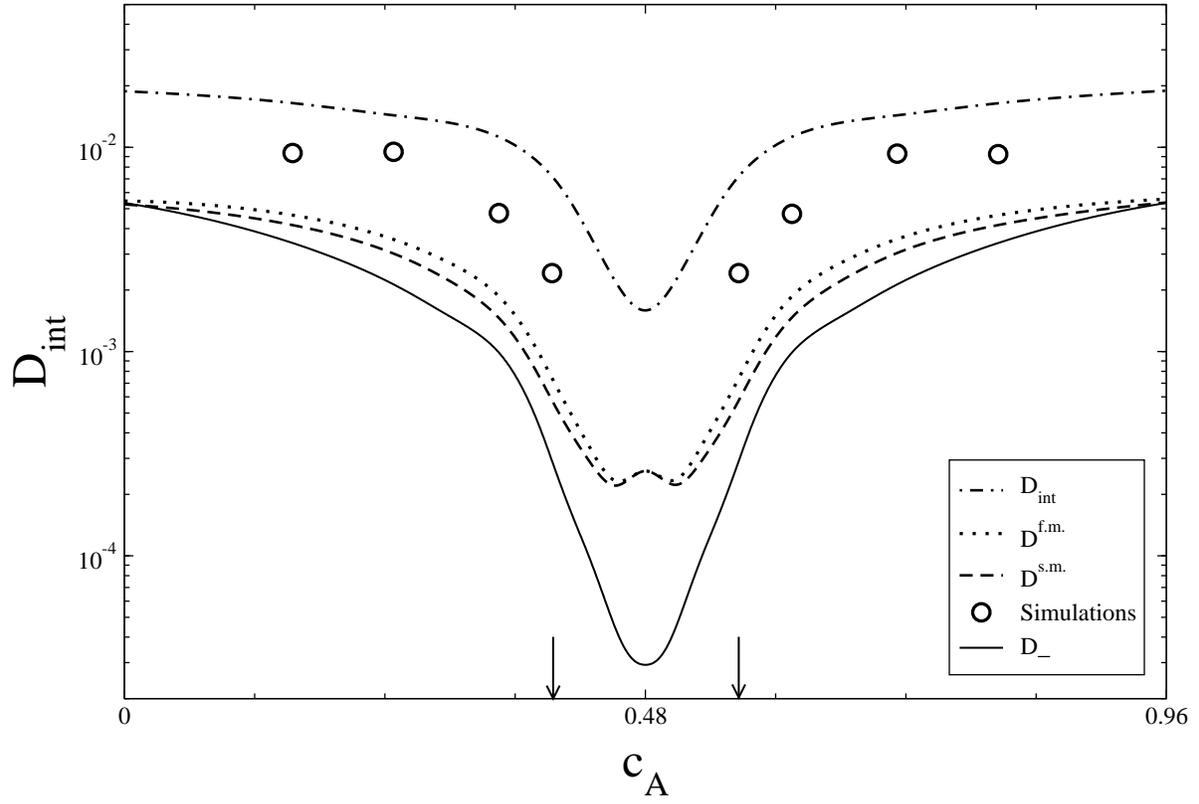}}
  \end{center}
  \caption{
      Plots of the interdiffusion coefficient  as a function of the 
      concentration $c_A$, for a temperature $T=0.6$. The jump rates are 
      $\Gamma_A = \Gamma_B = 1$. The arrows mark the corresponding critical 
      values of $c_A$ for this temperature.
      Simulation results are compared to different theoretical approaches,
      for a discussion see the text. 
       }   \label{fig15}
\end{figure}

\begin{figure}
  \begin{center}
     \scalebox{0.55}{\includegraphics{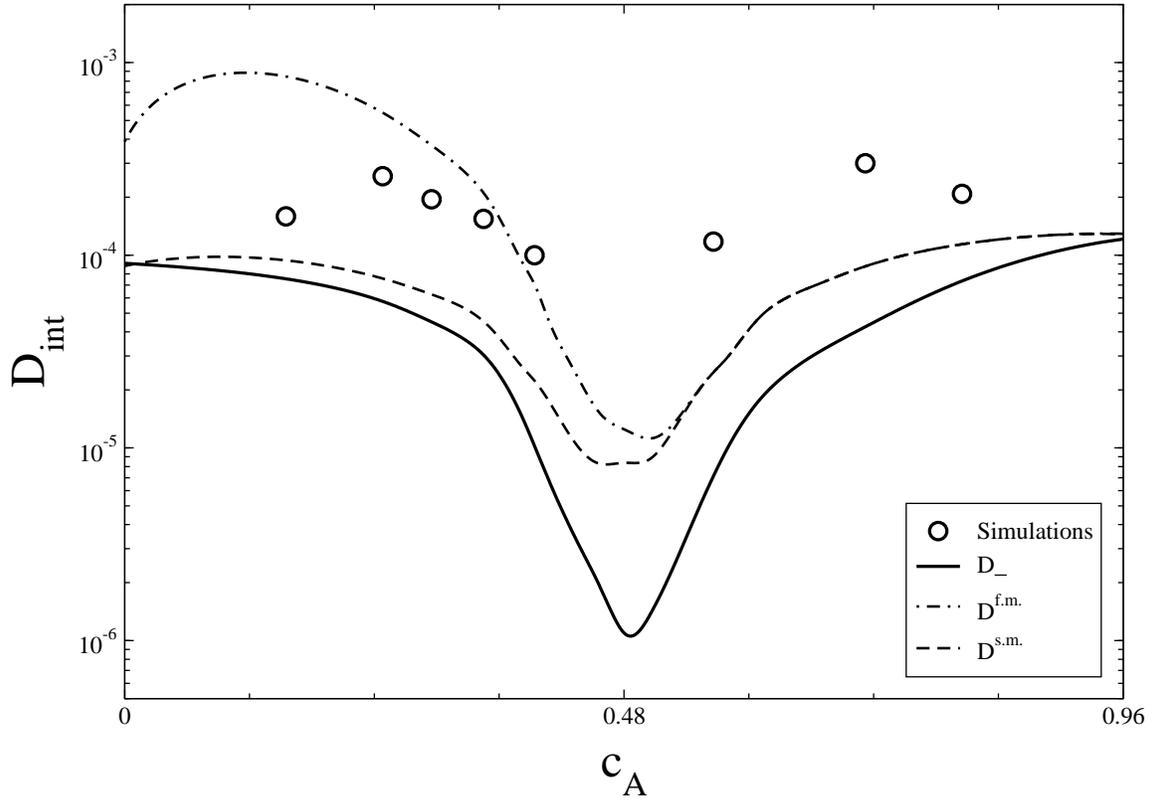}}
  \end{center}
  \caption{
      Plots of the interdiffusion coefficient as a function of the 
      concentration $c_A$, for a temperature $T=0.6$. The jump rates are 
      $\Gamma_A / \Gamma_B = 0.01$.
       }   \label{fig16}
\end{figure}

\begin{figure}
  \begin{center}
     \scalebox{0.55}{\includegraphics{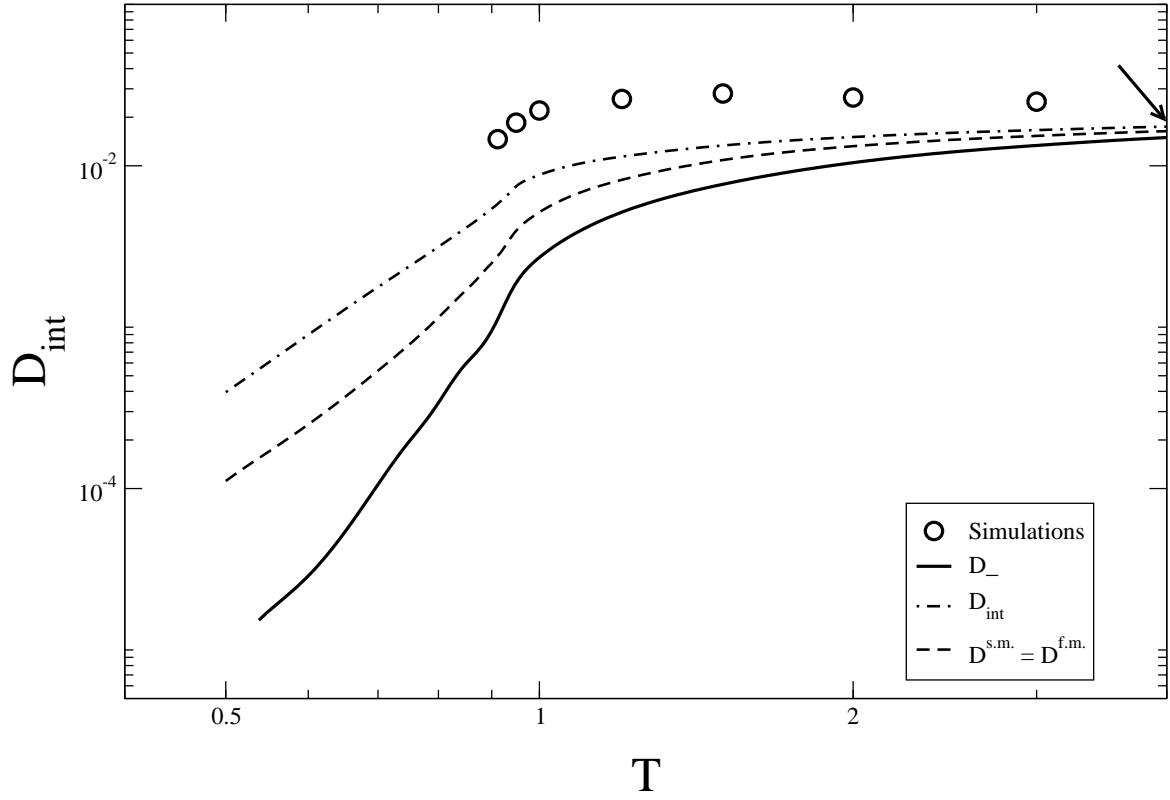}}
  \end{center}
  \caption{
    Logarithmic plot of the interdiffusion coefficient as a function
    of the temperature, for a concentration $c_A = c_B = 0.48$.
      The jump rates are $\Gamma_A = \Gamma_B = 1$. The arrow marks 
     the infinite temperature result, where all the quantities showed 
      in the plot coincide.
       }   \label{fig17}
\end{figure}

\begin{figure}
  \begin{center}
     \scalebox{0.55}{\includegraphics{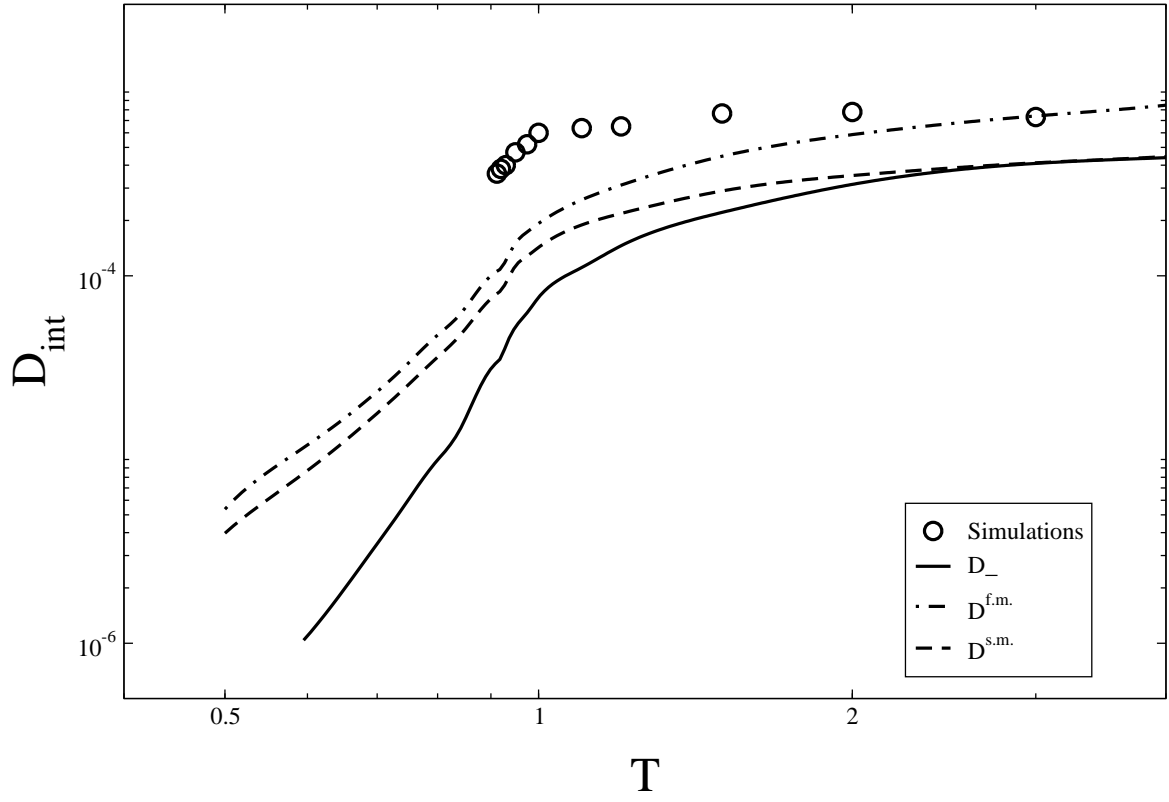}}
  \end{center}
  \caption{
       Logarithmic plot of the interdiffusion coefficient as a function
      of the temperature, for a concentration $c_A = c_B = 0.48$.
       The jump rates are $\Gamma_A / \Gamma_B = 0.01$. 
       }  \label{fig18}
\end{figure}

\end{document}